\documentclass[pre,twocolumn,byrevtex,superscriptaddress]{revtex4}
\usepackage{amsmath,epsfig,enumerate,graphicx}
\usepackage{verbatim}
\usepackage{amstext}
\usepackage{ifthen}
\newboolean{twocolswitch}
\newboolean{revtexswitch}

\newcommand{\Ai}[1]{#1}

\newcommand{\partialdiff}[2]{\frac{\partial #1}{\partial #2}}
\newcommand{\tpartialdiff}[2]{\partial #1/\partial #2}

\newcommand{\req}[1]{(\ref{#1})}

\newcommand{\etal}{\textit{et al.}}

\newcommand{\tempc}[1]{${#1}^{\circ}\mbox{C}$}

\newcommand{\Mmax}{M_{\mbox{\scriptsize{max}}}}
\newcommand{\Mmin}{M_{\mbox{\scriptsize{min}}}}
\setboolean{twocolswitch}{true}
\setboolean{revtexswitch}{true}

\begin{document}

\title{
Re-examination of the ``3/4-law'' of Metabolism
}

\author{
  \firstname{Peter Sheridan}
  \surname{Dodds}
  }
\thanks{Author to whom correspondence should be addressed}
\email{dodds@segovia.mit.edu}
\homepage{http://segovia.mit.edu/}
\affiliation{
        Department of Mathematics,
        Massachusetts Institute of Technology,
        Cambridge, MA 02139.
        }
\affiliation{
        Department of Earth, 
        Atmospheric and Planetary Sciences,
        Massachusetts Institute of Technology, 
        Cambridge, MA 02139.
        }

\author{
  \firstname{Daniel H.}
  \surname{Rothman}
  }
\email{dan@segovia.mit.edu}
\affiliation{
        Department of Earth, 
        Atmospheric and Planetary Sciences,
        Massachusetts Institute of Technology, 
        Cambridge, MA 02139.
        }

\author{
  \firstname{Joshua S.}
  \surname{Weitz}
  }
\email{jsweitz@segovia.mit.edu}
\affiliation{
        Department of Earth, 
        Atmospheric and Planetary Sciences,
        Massachusetts Institute of Technology, 
        Cambridge, MA 02139.
        }
\affiliation{
        Department of Physics,
        Massachusetts Institute of Technology,
        Cambridge, MA 02139.
        }

\date{\today}

\begin{abstract}
We examine the scaling law $B\propto M^{\alpha}$ which connects organismal 
metabolic rate $B$ with organismal mass $M$,
where $\alpha$ is commonly held to be $3/4$.
Since simple dimensional analysis suggests $\alpha=2/3$, 
we consider this to be a null hypothesis 
testable by empirical studies.
We re-analyze data sets for mammals and birds 
compiled by Heusner, Bennett and Harvey, Bartels, 
Hemmingsen, Brody, and Kleiber, 
and find little evidence for rejecting $\alpha=2/3$
in favor of $\alpha=3/4$.
For mammals, we find a possible breakdown in scaling for larger masses
reflected in a systematic increase in $\alpha$.
We also review theoretical justifications of $\alpha=3/4$ based on
dimensional analysis, nutrient-supply networks, and four-dimensional biology.
We find that present theories for $\alpha=3/4$ require assumptions
that render them unconvincing for rejecting the null hypothesis that $\alpha=2/3$.
\end{abstract}

\maketitle

\section{Introduction}
The ``$3/4$-law'' of metabolism states
that organismal basal metabolic rate, $B$,
is related to organismal mass, $M$, via
the power law~\citep{kleiber32,kleiber61,bonner83,calder84,schmidt-nielsen84,peters83}
\begin{equation}
  \label{eq:truemet.metabolic}
  B =  c M^{\alpha},
\end{equation}
where $\alpha$ is believed to be $3/4$.
The assumption that $\alpha=3/4$ is
relevant in medicine~\citep{mordenti86,anderson97,mahmood99},
nutrition~\citep{cunningham80,pike84,burger91},
and ecology~\citep{damuth81,lindstedt86,calder84,carbone99}, 
and has been the subject of a series of theoretical
debates~\citep{blaxter65,heusner82b,feldman82,economos83,prothero84a,feldman95}.
It has been oft quoted that quarter-power
scaling is ubiquitous in biology~\citep{calder84,west97}.
Such quarter-law scaling reinforces, and is reinforced by, 
the notion that basal metabolic rate scales like $B\propto M^{3/4}$. 

Nevertheless, the reasons, biological or otherwise, for why $\alpha=3/4$
have remained elusive and their elucidation stands as an open theoretical problem.
A recent surge of interest
in the subject, including our own,  
has been inspired by the elegant attempt 
of \ifthenelse{\boolean{revtexswitch}}{West, Brown and Enquist~\citep{west97}}{\citet*{west97}}
to link nutrient-supply networks to metabolic scaling.
This work suggests that a fundamental understanding of the
relationship between basal metabolism and body size
is within our grasp and that closer inspection of
both theory and data are duly warranted.

In this paper
we work from the null hypothesis that $\alpha=2/3$.
In a resting state, heat is 
predominantly lost through the surface of a body.
One then expects, from naive dimensional analysis, that
basal metabolism scales as surface area which scales as 
$V^{2/3}$ where $V$, volume, 
is proportional to $M$ presuming density is constant.  
This scaling of surface area with mass has found strong empirical 
support in organismic biology~\citep{hemmingsen60,schmidt-nielsen84,calder84,heusner87}.
Such a surface law of metabolism
was first expounded in the nineteenth century~\citep{rubner1883}.
Later observations of deviations from $\alpha=2/3$ eventually 
led to its replacement by $\alpha\simeq 0.72$--$0.73$ which
was then supplanted by the simpler $\alpha=3/4$~\citep{brody45,hemmingsen60,kleiber61}.
\ifthenelse{\boolean{revtexswitch}}
{
  The widespread agreement that $\alpha=3/4$
  is due largely to the formative influence of 
  Kleiber~\citep{kleiber32,kleiber61,schmidt-nielsen84note}
  and has been accepted and 
  used as a general rule for decades~\citep{blaxter65}.
}
{
  The widespread agreement that $\alpha=3/4$
  is due largely to the formative influence of 
  \citet{kleiber32,kleiber61}\footnote{
    Kleiber's motivation in part was to make calculations
    less cumbersome with a slide rule~\citep[p. 59]{schmidt-nielsen84}.    
    }
  and has been accepted and 
  used as a general rule for decades~\citep{blaxter65}.
}

We re-examine empirical data available for
metabolic rates of homoiotherms as well as carefully review both recent
and historical theoretical justifications for $\alpha=3/4$.
Our statistical analysis of data collated by 
\ifthenelse{\boolean{revtexswitch}}{Heusner~\citep{heusner91}}{\citet{heusner91}}
for 391 species of mammals and by \ifthenelse{\boolean{revtexswitch}}{Bennett and Harvey~\citep{bennett87}}{\citet{bennett87}}
for 398 species of birds shows that over considerable, but not all,
ranges of body size, the hypothesis $\alpha=2/3$ is not rejected by the available data.
We also review empirical studies by
\ifthenelse{\boolean{revtexswitch}}{Bartels~\citep{bartels82}}{\citet{bartels82}}, 
\ifthenelse{\boolean{revtexswitch}}{Hemmingsen~\citep{hemmingsen60}}{\citet{hemmingsen60}}, 
\ifthenelse{\boolean{revtexswitch}}{Brody~\citep{brody45}}{\citet{brody45}}, and 
\ifthenelse{\boolean{revtexswitch}}{Kleiber~\citep{kleiber32}}{\citet{kleiber32}}
and find the data, upon re-examination, to be supportive of our interpretations.
We then examine theoretical attempts to connect metabolic rate to mass.
These include approaches based on 
dimensional analysis~\citep{gunther82,economos82,gunther85,bonner83,heusner82a,feldman95},
four-dimensional biology~\citep{blum77,west99},
and
nutrient-supply networks~\citep{west97,banavar99}.
We find that none of these theories convincingly show that
$\alpha=3/4$, rather than $\alpha=2/3$, should be expected.

However, we do not suggest that the $3/4$-law 
should be replaced by a $2/3$-law of allometric scaling.  
We instead argue for a more general approach to the subject, 
using $\alpha=2/3$ as a null hypothesis which should be tested
by empiricists when considering $\alpha\neq 2/3$, and 
acknowledge
the possibility of deviations from simple scaling.

\section{Measuring the metabolic exponent}
\label{sec:truemet.data}

The history of metabolic scaling 
may be traced through a series of heavily cited empirical papers, 
some of which are composed of very few data points. 
In order to better understand the
scaling of metabolic rate, we work back in time, calculating $\alpha$ 
and deviations from uniform scaling for data from
\ifthenelse{\boolean{revtexswitch}}{Heusner~\citep{heusner91}}{\citet{heusner91}}, 
\ifthenelse{\boolean{revtexswitch}}{Bennett and Harvey~\citep{bennett87}}{\citet{bennett87}}, 
\ifthenelse{\boolean{revtexswitch}}{Bartels~\citep{bartels82}}{\citet{bartels82}}, 
\ifthenelse{\boolean{revtexswitch}}{Hemmingsen~\citep{hemmingsen60}}{\citet{hemmingsen60}}, 
\ifthenelse{\boolean{revtexswitch}}{Brody~\citep{brody45}}{\citet{brody45}}, and 
\ifthenelse{\boolean{revtexswitch}}{Kleiber~\citep{kleiber32}}{\citet{kleiber32}}. 
These papers represent some of the most influential, widely cited,
and often controversial papers in the field.
Our re-analysis of the data demonstrates that $\alpha=2/3$ should not be
rejected for mammals with mass less than approximately $10$--$35$ kg, 
and a similar analysis of
metabolic data for birds demonstrates $\alpha=2/3$ should not be rejected
for birds in general.

We have used the same methods to calculate $\alpha$
and its dependence on $M$ in all cases where
data is available.   Slopes and intercepts are determined
using standard linear regression in log-space taking $M$ to be the
independent variable.  The standard correlation
coefficient is denoted by $r$ while that 
obtained using the Spearman rank ordering~\citep{press92} is written as $r_s$.
When data is not available we have attempted to
classify the data sets in terms of the original calculations of $\alpha$ 
and its dependence on $M$.

\subsection{Heusner (1991)}

\begin{figure}[tbp!]
  \begin{center}
    \ifthenelse{\boolean{twocolswitch}}
    {
      \epsfig{file=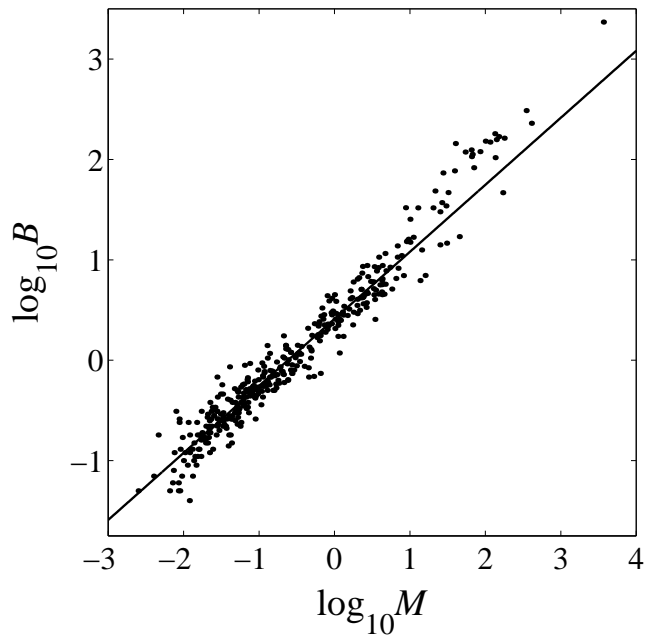,width=.48\textwidth}
      }
    {
      \epsfig{file=figheusner391_noname.eps,width=.7\textwidth}
      }
\ifthenelse{\boolean{revtexswitch}}
{
    \caption{
      Metabolic rate, $B$ (watts), as a function of mass, $M$ (kg), for 391 species of mammals.
      Data taken from Heusner~\citep{heusner91}.
      The straight line represents the best fit for the
      357 species with mass less than 10 kg where
      the $\alpha = 0.668 \pm 0.019$.
      The upward deviations for species with larger mass
      (see Table~\ref{tab:truemet.heusner391})
      may indicate a real biological difference but may also be due to the
      paucity of data.
      }
}
{
    \caption{
      Metabolic rate, $B$ (watts), as a function of mass, $M$ (kg), for 391 species of mammals.
      Data taken from \citet{heusner91}.
      The straight line represents the best fit for the
      357 species with mass less than 10 kg where
      the $\alpha = 0.668 \pm 0.019$.
      The upward deviations for species with larger mass
      (see Table~\ref{tab:truemet.heusner391})
      may indicate a real biological difference but may also be due to the
      paucity of data.
      }
}
    \label{fig:truemet.heusner391}
  \end{center}
\end{figure}

Data on basal metabolic rate for 391 mammalian species
compiled by
\ifthenelse{\boolean{revtexswitch}}{Heusner~\citep{heusner91}}{\citet{heusner91}}
is reproduced
in Figure~\ref{fig:truemet.heusner391}.
Heusner proposed that species could be separated
into two groups, one of
animals whose basal metabolism is normally
distributed about a regression line
and one of statistical outliers.  
Both groups were found by Heusner to satisfy a $2/3$-law for metabolism.

The results of simple regression analysis over
various mass ranges for Heusner's data
are shown in Table~\ref{tab:truemet.heusner391}.
We observe a break in scaling occurring
at around $M \simeq 10$ kg.
For those ranges with an upper mass $\Mmax \leq 10$ kg,
$\alpha=2/3$ appears to be robust.
Note that the data comprises 179 species of the order rodentia
ranging over more than three orders of magnitude of mass
from 0.007 kg to 26.4 kg.
On separating out these samples, we still find
$\alpha=0.675 \pm 0.025$ 
for the remaining species with $M \leq 10$ kg and
$\alpha=0.681 \pm 0.035$ for the rodentia species.

\begin{table}[tbp]
  \begin{center}
    \begin{tabular}{ccccccc}
      $\Mmax$ & $N$ & $\alpha$ & $\sigma(\alpha)$ & 0.05 & 0.01\\ \hline
      $0.01$    &  17 & 0.454 & 0.441 &  [-0.811,1.719] & [-1.294,2.202] \\
      $0.032$   &  81 & 0.790 & 0.093 &  [0.545,1.034] & [0.473,1.106] \\
      $0.1$    & 167 & 0.678 & 0.038 &  [0.578,0.778] & [0.550,0.806] \\
      $1$    & 276 & 0.662 & 0.016 &  [0.620,0.704] & [0.608,0.716] \\
      $10$    & 357 & 0.668 & 0.010 &  [0.643,0.693] & [0.636,0.700] \\
      $32$    & 371 & 0.675 & 0.009 &  [0.651,0.698] & [0.645,0.705] \\
      $100$   & 381 & 0.698 & 0.009 &  [0.675,0.720] & [0.668,0.727] \\
      $1000$   & 390 & 0.707 & 0.008 &  [0.686,0.728] & [0.680,0.734] \\
      $3670$   & 391 & 0.710 & 0.008 &  [0.689,0.731] & [0.684,0.737] \\
    \end{tabular}
    \caption{
      The exponent $\alpha$ measured for varying ranges of mass, $M < \Mmax$ (kg),
      for Heusner's data~\citep{heusner91}.  For each mass range,
      $N$ is the sample number and
      the errors, $\sigma(\alpha)$, are for two standard deviations.  
      The 95\% and 99\% confidence intervals of $\alpha$ are listed in the 
      last two columns.
      For small mammals ($M \leq 0.01$ kg, 17 species) 
      a large error is apparent but for increasing $\Mmax$, $\alpha$ 
      centers around $2/3$.  A gradual upwards drift in $\alpha$ is evident 
      for $\Mmax > 10$ kg.
      }
    \label{tab:truemet.heusner391}
  \end{center}
\end{table}

Upon addition of mammals with mass exceeding $10$ kg, 
the exponent steadily increases.  
Given the small number of samples of large mammals, 
one can only speculate on
the reason for this possible deviation.
Primarily, it may indicate a real upwards deviation from scaling, 
with larger organisms actually having greater metabolic rates
than predicted by $\alpha=2/3$~\citep{bartels82,economos83,heusner91a}.  
Larger organisms
are reported to scale allometrically in form
so such a deviation may be a result of changes 
in body shape and hence surface area~\citep{bonner83,calder84}.
Support for this notion comes from
\ifthenelse{\boolean{revtexswitch}}{Economos~\citep{economos82}}{\citet{economos82}}
who finds the relationship between
mammalian head-and-body length and mass is better fit
by two scaling laws rather than one.  
He identifies 20 kg as a breakpoint,
which is in accord with our findings here,
suggesting that geometric scaling holds below this mass while
allometric quarter-power scaling holds above.
The upper scaling observed by Economos could also be viewed as part of a gradual
deviation from geometric scaling.

The upwards shift of metabolic rates for larger mammals
could otherwise point to problems of measurement 
(note the corrections for 
elephants in 
\ifthenelse{\boolean{revtexswitch}}{Brody's data~\citep{brody45}}{\citet{brody45}}),
an evolutionary advantage related to larger brain
sizes~\citep{jungers85,allman99}, or the lack of competition 
for ecological niches for large mammals creating a distinction
with smaller mammals.

\subsection{Bennett and Harvey (1987)}
\begin{figure}[htbp]
  \begin{center}
    \ifthenelse{\boolean{twocolswitch}}
    {
      \epsfig{file=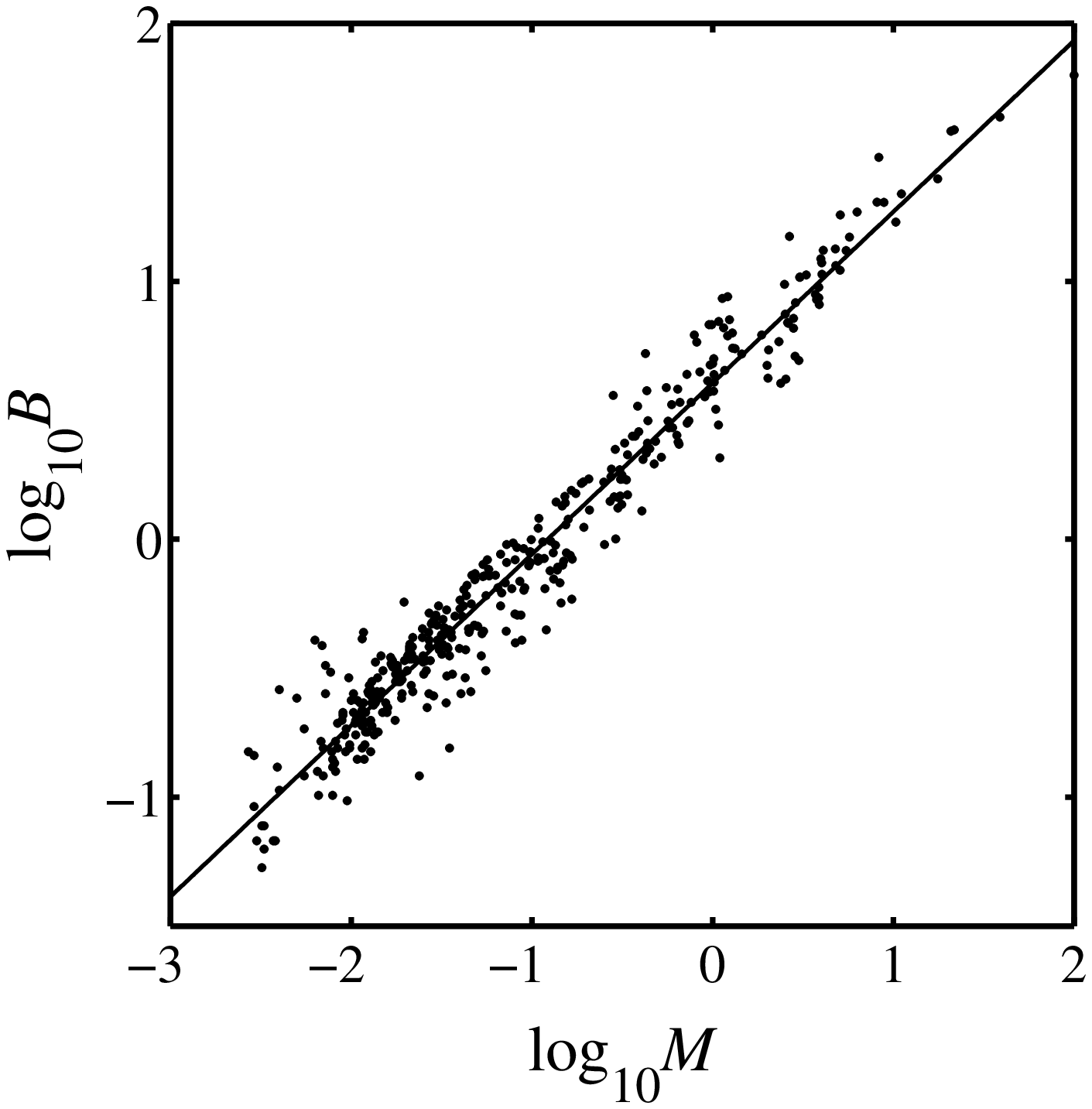,width=.48\textwidth}
      }
    {
      \epsfig{file=figbennettbirds398_noname.eps,width=.7\textwidth}
      }
\ifthenelse{\boolean{revtexswitch}}
{
    \caption{
      Metabolic rate, $B$ (watts), as a function of mass, $M$ (kg), for 
      398 species of birds taken from
      Bennett and Harvey~\citep{bennett87}.  
      The straight line is the
      result of a regression analysis which
      gives $\alpha = 0.666 \pm 0.014$.
      }
}
{
    \caption{
      Metabolic rate, $B$ (watts), as a function of mass, $M$ (kg), for 
      398 species of birds taken from
      \citet{bennett87}.  
      The straight line is the
      result of a regression analysis which
      gives $\alpha = 0.666 \pm 0.014$.
      }
}
    \label{fig:truemet.bennettbirds}
  \end{center}
\end{figure}

Birds show strong support for not rejecting the
null hypothesis $\alpha=2/3$.
\ifthenelse{\boolean{revtexswitch}}
{
  Figure~\ref{fig:truemet.bennettbirds} shows metabolic data 
  for 398 distinct bird species taken from
  Bennett and Harvey~\citep{bennett87,bennett87note}.
  }
{
  Figure~\ref{fig:truemet.bennettbirds} shows metabolic data 
  for 398 distinct bird species taken from
  \citet{bennett87}\footnote{
    Following \citet{bennett87}, 
    we take one sample for each species of bird selecting
    those with lowest mass-specific resting metabolic rate.
    Note that we also include organisms that Bennett and Harvey 
    state were measured during 
    their active cycle whereas Bennett and Harvey do not.
    The use of other selection criteria does not greatly affect
    the results we present here.}.
  }
We find here that $\alpha=0.666\pm 0.013$ in agreement
with Bennett and Harvey's calculations.
\ifthenelse{\boolean{revtexswitch}}{Lasiewski and Dawson~\citep{lasiewski67}}{\citet{lasiewski67}}
similarly found that
$\alpha=0.668$ for a smaller set of data.
Attempts to reconcile the $3/4$-law with these measurements have centered around
the division of birds into passerine (perching birds) and non-passerine species
(non-perching birds).
\Ai{Lasiewski} and \Ai{Dawson},
for example, found exponents $0.724$
and $0.723$ for passerine and non-passerine species respectively.
Though this is not an arbitrary division (core temperatures
are thought to differ by 1--\tempc{2}),
later work by 
\ifthenelse{\boolean{revtexswitch}}{Kendeigh~\citep{kendeigh77}}{\citet{kendeigh77}}
finds exponents ranging from
$0.668$--$0.735$ when passerines and non-passerines are
grouped according to different measurement conditions (winter vs.\ summer, etc.).

Similar distinctions between intra- and inter-species
scaling have been raised in the study of metabolic
scaling for mammals where it has been suggested that 
$\alpha=2/3$ for single species comparisons and $\alpha=3/4$
holds across differing species~\citep{schmidt-nielsen84,heusner82b,bonner83}.
\ifthenelse{\boolean{revtexswitch}}{Bennett and Harvey~\citep{bennett87}}{\citet{bennett87}}
also found that $\alpha$
depends on the level of taxonomic detail one is investigating.
It remains unclear whether such subdivisions 
reflect relevant biological distinctions
or underlying correlations in the choice of taxonomic levels. 

\subsection{Bartels (1982)}
\ifthenelse{\boolean{revtexswitch}}{Bartels~\citep{bartels82}}{\citet{bartels82}} 
analyzes a set of approximately 85 mammalian species.
Although data is not provided in the paper, a summary of his 
results can be found in Table~\ref{tab:truemet.bartels}.
Bartels finds $\alpha=0.66$ (no error estimate is given) for mammals with mass between
$2.4\times10^{-3}$ kg and $3800$ kg and concludes that the deviation
from the expected $3/4$ scaling is due to the variations
in metabolic rates of small animals.  This lends further weight
to our conjecture that there may be a mass dependence
of metabolic rate scaling.

\begin{table}[tbp]
  \begin{center}
    \begin{tabular}{ccccc}
      $\Mmin$       & $\Mmax$    &  $N$              & $\alpha$      & $r$ \\ \hline
      $2.4\times10^{-3}$ & 3800  & $\simeq 85$      & 0.66          & 0.99 \\
      $2.4\times10^{-3}$ & 0.26  & $\simeq 40$      & 0.42          & 0.76 \\
      0.26               & 3800  & $\simeq 45$      & 0.76          & 0.99 \\
    \end{tabular}
    \ifthenelse{\boolean{revtexswitch}}
    {
    \caption{
      Exponents measured for varying ranges of mass (kg), $\Mmin\leq M\leq \Mmax$ according
      to Bartels~\citep{bartels82}.  Here $N$ is the sample
      number and $r$ is the correlation coefficient.
      }
    }
    {
    \caption{
      Exponents measured for varying ranges of mass (kg), $\Mmin\leq M\leq \Mmax$ according
      to \citet{bartels82}.  Here $N$ is the sample
      number and $r$ is the correlation coefficient.
      }
    }
    \label{tab:truemet.bartels}
  \end{center}
\end{table}

\subsection{Hemmingsen (1960)}
Hemmingsen's data set~\citep{hemmingsen60} for mammals comprises
15 data points with masses between 0.01 kg and 3500 kg.
Most of his data is derived
from earlier work by Brody.  
He states that the data is well modeled by a power law
with $\alpha=0.73$.  
To reach this conclusion 
he does not compute the power law of best fit, but rather,
the ``straight line\ldots was chosen corresponding
to [$\alpha=$]$0.73$, as established by Kleiber and also by
Brody.''

Hemmingsen also finds that a $3/4$-law holds
for unicellular organisms.  
Hemmingsen's work has been cited extensively
in support of the claim that the
$3/4$-law is a universal biological 
phenomenon~\citep{peters83,schmidt-nielsen84,calder84,west97}.
A careful re-examination of Hemmingsen's work by 
\ifthenelse{\boolean{revtexswitch}}{Prothero~\citep{prothero86b}}{\citet{prothero86b}}
showed that $\alpha$ can range from approximately
0.60 to 0.75 depending on which unicellular organisms
are included in the regression.
In addition to these questions about scaling for unicellular life, 
\ifthenelse{\boolean{revtexswitch}}{Patterson~\citep{patterson92b,patterson92a}}{\citet{patterson92b,patterson92a}}
has theoretically 
shown for aquatic inverterbrates and algae that the scaling exponent can
range from 0.31 to 1.00 depending on the mass transfer mechanisms involved.  
We agree with Prothero's conclusions that
``a three-quarters power rule expressing energy metabolism
as a function of size in unicellular organisms generally is not
at all persuasive''~\citep{prothero86b}.

\subsection{Brody (1945)}
One of the most influential works on scaling and metabolism
is that of 
\ifthenelse{\boolean{revtexswitch}}{Brody~\citep{brody45}}{\citet{brody45}}.  Indeed, the scaling law for
metabolism is sometimes cited as the Brody-Kleiber law.
Brody compiles a list of metabolic rates for 67 mammals.
The complete data set yields $\alpha=0.73\pm 0.01$.  
However, on inspection, one makes the surprising observation
that 32 data points are artificial in that most of these are calculated
using previously determined empirical equations while
a few have been corrected to account for variations in animal activity.
Using the remaining set of 35 animals we 
nevertheless find
$\alpha=0.72\pm 0.02$. 
It is also important to note that Brody's research
was done before the widespread use of electrocardiography and,
as pointed out by 
\ifthenelse{\boolean{revtexswitch}}{Kinnear and Brown~\citep{kinnear67}}{\citet{kinnear67}}, 
Brody's data may contain overestimations of basal metabolic rate.

We re-analyze Brody's raw, uncorrected data for mammals
over different mass ranges as shown in Table~\ref{tab:truemet.brody}.
Again, an increase in $\alpha$ is observed for ranges of larger masses.
This is consistent with the
results from Heusner's and Bartel's data which
suggest a deviation from perfect scaling with increase in mass.
Furthermore, it is evident
that $\alpha=0.72$ as calculated by regression on the full data set is misleading.
We reiterate that we are not suggesting that there is any
robust scaling law for large masses.  The results of the
regression analysis merely suggest a dependence
of $\alpha$ on the mass ranges being considered and that
a strict power law may not be appropriate.

\begin{table}[tbp]
  \begin{center}
    \begin{tabular}{ccccc}
      $\Mmin$  & $\Mmax$  & $N$  & $\alpha$      & $r_s$  \\ \hline
      0.016    &  1  & 19      & $0.673 \pm 0.061$ & 0.91  \\
      0.016    & 10  & 26      & $0.709 \pm 0.020$ & 0.96 \\
      10       & 922 & 9       & $0.760 \pm 0.085$ & 0.95 \\
      0.016    & 922 & 35      & $0.718 \pm 0.022$ & 0.98 \\
    \end{tabular}
    \caption{
      Results of regression on Brody's data~\citep{brody45}
      over different mass intervals, $\Mmin < M < \Mmax$.
      An increase in $\alpha$
      occurs for ranges over larger masses.
      Here, $N$ is the number of data points
      and $r_s$ is the Spearman correlation coefficient.
      }
    \label{tab:truemet.brody}
  \end{center}
\end{table}

\subsection{Kleiber (1932)}
In his now-famous paper on metabolic rate, Kleiber 
analyzed 13 species of mammals with average 
mass ranging from 0.15 to 679 kg~\citep{kleiber32}.  
We find the scaling exponent for the data to be $0.738\pm 0.016$.
Again we consider the possibility of a crossover and separate the data 
into a set of 5 species with $M<10$ kg
and 8 species with $M>10$ kg.  
For $M<10$ kg, $\alpha=0.667 \pm 0.043$
while for $M>10$ kg, $\alpha=0.754\pm 0.048$.  
These results are again consistent with our assertion of a mass-dependent
$\alpha$.
Nevertheless, it is important to
remain mindful of the relative paucity of data in these
influential works.

\section{Fluctuations about scaling}
The next logical step after measuring the metabolic
exponent and systematic deviations thereof 
is to consider fluctuations about the mean.
This is seldom done with power law measurements~\citep{dodds2000ua}
and researchers concerned with the predictive power of a scaling 
law for metabolic rate 
have often pointed to organisms that deviate from predictions
as being either problematic or 
different~\citep{brody45,bartels82,schmidt-nielsen84,heusner91}.
We take the view that fluctuations are to be expected
and quantified appropriately.

We thus generalize the relation $B = cM^\alpha$
by considering
$P(B\, |M)$, the conditional probability density
of measuring a metabolic rate, $B$, given a mass, $M$,
\begin{equation}
  \label{eq:truemet.metascaling}
  P(B\, |M) = M^{-\alpha} f(B/M^{\alpha})
\end{equation}
where the leading factor of $M^{-\alpha}$ is for normalization
and $\int_{0}^\infty f(x) dx = 1$.

Our null hypothesis is that fluctuations are Gaussian in logarithmic space,
i.e., $f$ is a lognormal distribution.  
Gaussian fluctuations are typically assumed in statistical inferences
made using regression analysis~\citep{degroot1975}.  
Demonstrating that $f$ is not inconsistent with a normal distribution
will therefore allow us to use certain hypothesis tests in the following section.

If equation~\req{eq:truemet.metascaling} is correct then the sampled 
data can be rescaled accordingly to reconstruct
$f$, the scaling function. 
To do so, one must first determine $\alpha$.
We suggest the most appropriate estimate of $\alpha$
corresponds to the case when the residuals about the best fit
power law are uncorrelated with regards to body mass.   
This is similar to techniques used in the analysis of partial residuals~\citep{hastie87}
and we make use of it later.
We obtain residuals for the range $0.5 \leq \alpha \leq 1.5$
where the prefactor $c$ of $B = cM^\alpha$ is determined via least squares.
The Spearman correlation coefficient $r_s$ is then determined
for the residuals and recorded as a function of $\alpha$.
We then take the value of $\alpha$ for which
$r_s=0$ as the most likely underlying scaling exponent.

We find $r_s=0$ when
$\alpha\simeq 0.665$ for mammals using Heusner's data with $M \leq 10$ kg
and $r_s \simeq -0.42$ when $\alpha=3/4$.
For the entire range of masses in bird data of Bennett and Harvey,
$r_s=0$ when $\alpha\simeq 0.671$.

\begin{figure}[tbp]
  \begin{center}
    \ifthenelse{\boolean{twocolswitch}}
    {
      \epsfig{file=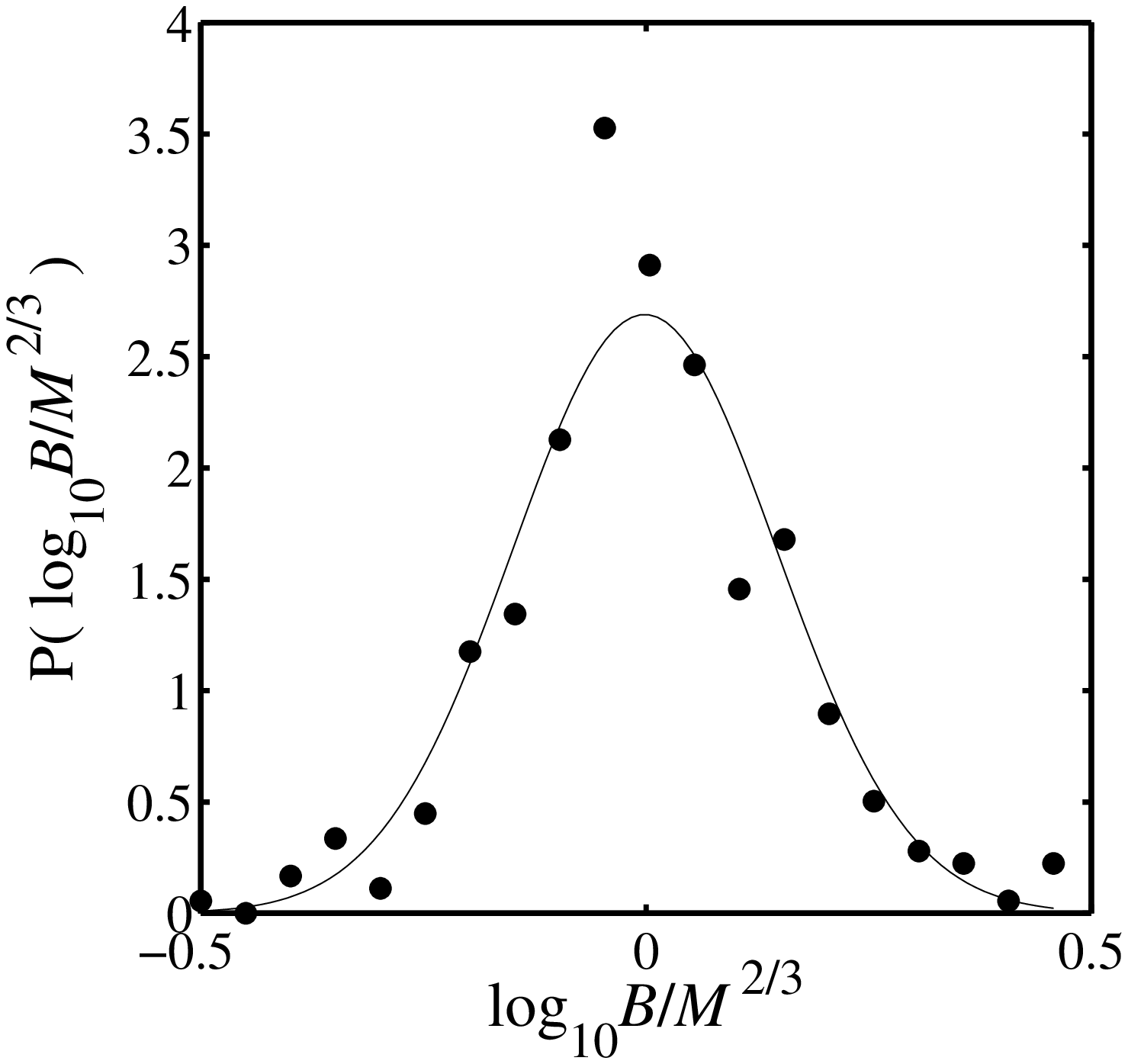,width=.48\textwidth}
      }
    {
      \epsfig{file=figmetascalingfn2_noname.eps,width=.7\textwidth}
      }
    \caption{
      Fluctuations in metabolic rate for mammals
      with $M \leq 10$ kg taken from Heusner's data set~\citep{heusner91}.
      The scaling function $f$ 
      (see equation~\req{eq:truemet.metascaling}) is
      fitted with a lognormal distribution.
      Values of $B$ have been renormalized as $B/M^{2/3}$
      and partitioned into 20 bins.
      }
    \label{fig:truemet.metascalingfn}
  \end{center}
\end{figure}

With these results in hand, we extract $f$ for mammals and birds,
the results for mammals being shown in Figure~\ref{fig:truemet.metascalingfn}.
We find the form of $f$ agrees qualitatively with a lognormal.
In order to quantify the quality of this agreement we employ
the Kolmogorov-Smirnov test~\citep{degroot1975},
a non-parametric test which gives a significance probability
($p$-value) for whether
or not a sample comes from a hypothesized distribution.
Not having a hypothesis for the value of the standard
deviation, we take two approaches to deal with this problem.
Asserting the measured sample standard deviation $\sigma$
to be that of the underlying normal distribution,
we calculate the corresponding significance probability, $p$.
Alternatively, an estimate of $\sigma$, $\sigma^{\ast}$, may be obtained
by finding the value of $\sigma$ which maximizes $p$ such that
$p(\sigma^{\ast})=p^{\ast}$.
Results for both calculations are found in Table~\ref{tab:truemet.residuals}.
All $p$-values are above 0.01, i.e., none show very significant
deviations.  Additionally, the $p$-value for only the case of the birds
using $\sigma$ estimated from the data falls below 0.05 indicating
its departure is significant, but this is balanced by the high $p$-value
found by the maximizing procedure.

Thus, we suggest the data supports the
simple hypothesis of lognormal fluctuations
around a scaling law with $\alpha \simeq 2/3$.

\begin{table}[tbp]
  \begin{center}
    \begin{tabular}{lccccc}
      & range  & $\sigma$ &    $p$  & $\sigma^{\ast}$ & $p^{\ast}$ \\ \hline
      mammals & $M < 1$  & 0.153    & 0.232   & 0.120 & 0.307 \\
      mammals & $M < 10$ & 0.153    & 0.093   & 0.120 & 0.135 \\
      birds   & all    & 0.132    & 0.032   & 0.115 & 0.573 \\ 
    \end{tabular}
    \caption{
      Results from Kolmogorov-Smirnov tests for the underlying distributions
      of fluctuations around pure scaling for both mammals~\citep{heusner91} and 
      birds~\citep{bennett87}.
      The distribution is assumed to be lognormal, i.e.,
      normal in logarithmic coordinates.
      The standard deviations $\sigma$ are calculated directly from the residuals themselves
      and determine a level of significance $p$.
      The $\sigma^{\ast}$ correspond to $p^{\ast}$, the maximum
      $p$-value possible. 
      }
    \label{tab:truemet.residuals}
  \end{center}
\end{table}

\section{Hypothesis tests}
We now construct two types of hypothesis tests
to determine whether or not
$\alpha=2/3$ or $\alpha=3/4$ should be rejected by the available data.
The first test is the standard method of testing
the results of a linear regression against a presumed slope.
The second is a natural extension
of examining fluctuations about a linear fit as per
the previous section.
By analyzing the correlations of the residuals from the best
fit line we are able to quantitatively determine which
values of $\alpha$ are compatible with the data.  
In both tests, we reject an hypothesis when $p<0.01$.

\subsection{Comparison to a fixed $\alpha$}
For a given set of $N$ measurements for both mass, $M_i$, and
metabolic rate, $B_i$, 
we pose the following hypotheses: 
\begin{eqnarray}
  \label{eq:truemet.H0H1}
  H_0: \alpha & = & \alpha', \\
  H_1: \alpha & \neq & \alpha'.
\end{eqnarray}
We test the null hypothesis, $H_0$, in the specific cases
$\alpha'=2/3$ and $\alpha'=3/4$ for data from
\ifthenelse{\boolean{revtexswitch}}{Kleiber~\citep{kleiber32}}{\citet{kleiber32}},
\ifthenelse{\boolean{revtexswitch}}{Brody~\citep{brody45}}{\citet{brody45}}, 
\ifthenelse{\boolean{revtexswitch}}{Bennett and Harvey~\citep{bennett87}}{\citet{bennett87}},
and \ifthenelse{\boolean{revtexswitch}}{Heusner~\citep{heusner91}}{\citet{heusner91}},
over various mass ranges.  
Here, the $p$-value represents the probability that,
given two variables linearly related with slope $\alpha'$
and subject to Gaussian fluctuations, a data set formed 
with $N$ samples would have a measured slope
$\alpha$ differing at least by $|\alpha-\alpha'|$ from 
$\alpha'$~\citep{degroot1975}.
For a null hypothesis
with $\alpha=\alpha'$, we write the $p$-value
as $p_{\alpha'}$, e.g., $p_{3/4}$.

For mammals with $M\leq 10$ kg, the
results of the hypothesis test are contained in
Table~\ref{tab:truemet.nullhypoth1}.
The null hypothesis that $\alpha=3/4$ is rejected for both Brody
and Heusner's data and should not be rejected
in the case of Kleiber. 
The alternative null hypothesis that
$\alpha=2/3$ is not rejected for both Heusner and Kleiber and
rejected in the case of Brody.  Again, divisions into mass ranges are somewhat
arbitrary and are chosen to help demonstrate the mass-dependence of $\alpha$.
For example, for mammals with $M<1$ kg, 
Brody's data implies we should not reject the hypothesis that $\alpha=2/3$.

\begin{table}[htbp]
  \begin{center}
    \begin{tabular}{cccccc}
              & $N$ & $\alpha$ & $\sigma(\alpha)$ & $p_{2/3}$ & $p_{3/4}$ \\ \hline
      Kleiber        &   5 & 0.667 & 0.016 & 0.99 & 0.088 \\
      Brody        &  26 & 0.709 & 0.020 & $<10^{-3}$ & $<10^{-3}$ \\
      Heusner        & 357 & 0.668 & 0.010 &  0.91 &   $<10^{-15}$ \\
    \end{tabular}
    \caption{
      Hypothesis test based on standard comparison between slopes that 
      $\alpha=2/3$ and $\alpha=3/4$ for mammals with $M \leq 10$ kg.  
      Here, $\alpha$ is the measured exponent, $\sigma$
      is the standard error, and the $p$-values $p_{2/3}$ and $p_{3/4}$
      for the hypothesis $\alpha=2/3$ and
      $\alpha=3/4$ are listed in the last two columns.
      }
    \label{tab:truemet.nullhypoth1}
  \end{center}
\end{table}

Table~\ref{tab:truemet.nullhypoth2}
details results for mammals with $M\geq 10$ kg.
In the smaller data sets of
Kleiber and Brody the hypothesis
that $\alpha=3/4$ is not rejected while for the larger data set
of Heusner, $\alpha=3/4$ is rejected.  In all cases
the hypothesis that $\alpha=2/3$ for large mammals is rejected.
Even though Brody and Kleiber's data sets are consistent with an exponent $\alpha>3/4$, the
relative lack of metabolic measurements on large mammals 
and the strong rejection by Heusner's larger sample prevents
us from drawing definitive conclusions about the particular value, if any, of $\alpha$
for $M \geq 10$ kg.

\begin{table}[htbp]
  \begin{center}
    \begin{tabular}{cccccc}
              & $N$ & $\alpha$ & $\sigma(\alpha)$ & $p_{2/3}$ & $p_{3/4}$ \\ \hline
      Kleiber        &   8 & 0.754 & 0.021 & $<10^{-4}$ & 0.66 \\
      Brody          &   9 & 0.760 & 0.038 & $<10^{-3}$ & 0.56 \\
      Heusner        &  34 & 0.877 & 0.088 & $< 10^{-12}$ & $<10^{-7}$ \\
    \end{tabular}
    \caption{Hypothesis test based on standard comparison between slopes that 
      $\alpha=2/3$ and $\alpha=3/4$ for mammals with $M \geq 10$ kg.  
      See Table~\ref{tab:truemet.nullhypoth1}
      for the definition of all quantities.}
    \label{tab:truemet.nullhypoth2}
  \end{center}
\end{table}

When all mass ranges are considered for both birds and mammals
the hypothesis test (see Table~\ref{tab:truemet.nullhypoth3})
demonstrates that both $\alpha=2/3$
and $\alpha=3/4$ are rejected based on the empirical
data on mammals, while $\alpha=2/3$ is not rejected and $\alpha=3/4$ is rejected
based on the empirical data on birds.  
In summary, we find that a single exponent 
may be appropriate for rough estimates but, from a
statistical point of view, it appears that no single exponent explains
the data on metabolic scaling for mammals.

\begin{table}[htbp]
  \begin{center}
    \begin{tabular}{cccccc}
             & $N$ & $\alpha$ & $\sigma(\alpha)$ & $p_{2/3}$ & $p_{3/4}$ \\ \hline
      Kleiber        &  13 & 0.738 & 0.007 & $<10^{-6}$ & 0.11 \\
      Brody        &  35 & 0.718 & 0.011 & $<10^{-4}$ & $<10^{-2}$ \\
      Heusner        & 391 & 0.710 & 0.008 & $<10^{-6}$ & $<10^{-5}$ \\
      Bennett and Harvey        & 398 & 0.664 & 0.007 & 0.69 &     $<10^{-15}$ \\
    \end{tabular}
    \caption{Hypothesis test based on standard comparison between slopes that 
      $\alpha=2/3$ and $\alpha=3/4$ for birds and mammals over their entire mass range.
      See Table~\ref{tab:truemet.nullhypoth1}
      for the definition of all quantities.}
    \label{tab:truemet.nullhypoth3}
  \end{center}
\end{table}

\subsection{Analysis of residuals}
As per our discussion of fluctuations, 
a sensitive test of a null hypothesis is to check
the rank-correlation coefficient of the residuals.
In order to test the hypothesis, $\alpha=\alpha'$,
we pose the following hypotheses:
\begin{eqnarray}
  \label{eq:truemet.H0H1residuals}
  H_0: r_{s,\alpha'}(z_i,M_i) & =  & 0, \\
  H_1: r_{s,\alpha'}(z_i,M_i) & \neq  &0.
\end{eqnarray}
where the $z_i$ are the residuals.
The hypothesis $H_0$ means that if the residuals for the
power law $B = c M^{\alpha'}$ are uncorrelated with $M$
then $\alpha'$ could be the underlying exponent.  
The alternative hypothesis $H_1$
means that the residual correlations are significant
and the null hypothesis should be rejected.
The $p$-values represent the probability
that the magnitude of the correlation of the residuals, $|r_{s,\alpha'}(z_i,M_i)|$,
would be at least its value as expected for samples taken
from randomly generated numbers.

\begin{figure}[tbp!]
  \begin{center}
    \ifthenelse{\boolean{twocolswitch}}
    {
      \epsfig{file=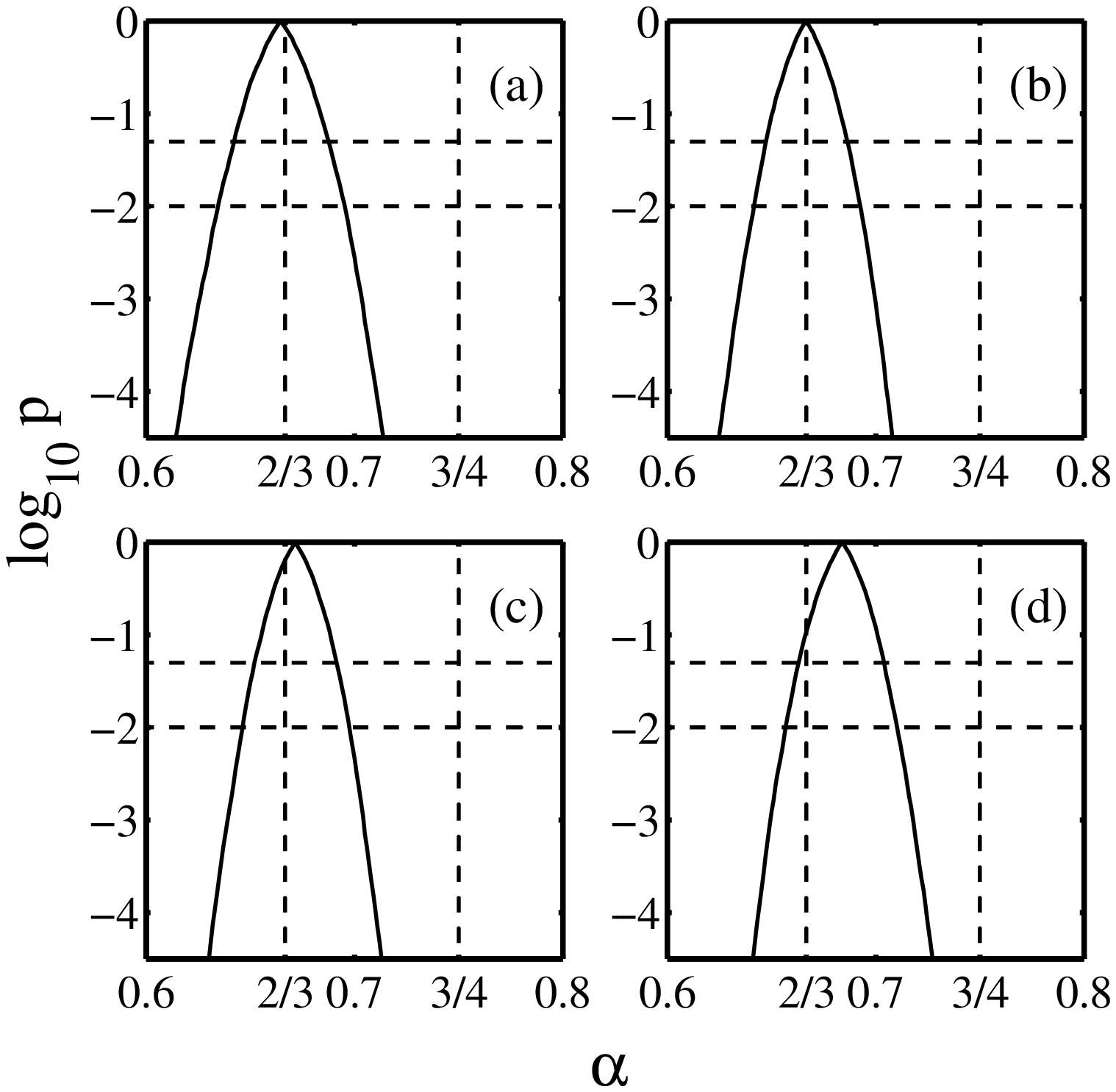,width=0.48\textwidth}
      }
    {
      \epsfig{file=figmammals_pv_log_noname.eps,width=0.75\textwidth}
      }
    \ifthenelse{\boolean{revtexswitch}}
{
    \caption{Test of the null hypothesis $H_0: r_{s,\alpha'}(z_i,M_i) =  0$
      based on mammalian data from Heusner~\citep{heusner91}
      (see equation~\req{eq:truemet.H0H1residuals}).
      Shown are plots of $p(\alpha)$ for differing mass ranges.
      In all plots the two dashed horizontal lines correspond to $p=0.05$ and $p=0.01$.  
      The individual plots correspond
      to the following ranges: (a) $M<3.2$ kg, (b) $M<10$ kg, (c) $M<32$ kg, (d) all mammals.
      For all mass ranges considered, $p_{2/3}>0.05$ and $p_{3/4} \ll 10^{-4}$.  
      }
}
{
    \caption{Test of the null hypothesis $H_0: r_{s,\alpha'}(z_i,M_i) =  0$
      based on mammalian data from \citet{heusner91}
      (see equation~\req{eq:truemet.H0H1residuals}).
      Shown are plots of $p(\alpha)$ for differing mass ranges.
      In all plots the two dashed horizontal lines correspond to $p=0.05$ and $p=0.01$.  
      The individual plots correspond
      to the following ranges: (a) $M<3.2$ kg, (b) $M<10$ kg, (c) $M<32$ kg, (d) all mammals.
      For all mass ranges considered, $p_{2/3}>0.05$ and $p_{3/4} \ll 10^{-4}$.  
      }
}
    \label{fig:truemet.heusner_resid_test}
  \end{center}
\end{figure}

\begin{figure}[tbp!]
  \begin{center}
    \ifthenelse{\boolean{twocolswitch}}
    {
      \epsfig{file=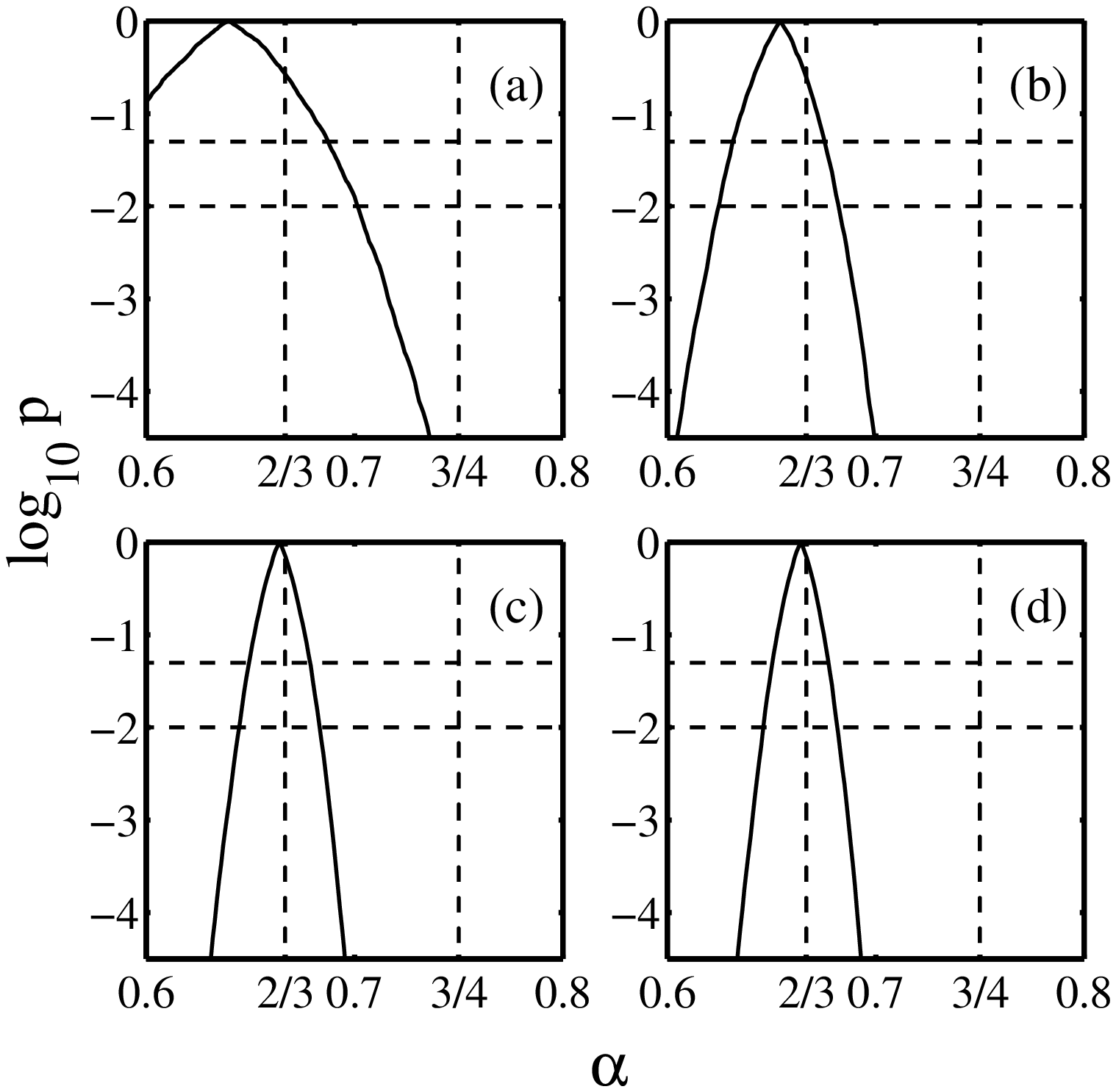,width=0.48\textwidth}
      }
    {
      \epsfig{file=figbirds_pv_log_noname.eps,width=0.75\textwidth}
      }
\ifthenelse{\boolean{revtexswitch}}
{
    \caption{Test of the null hypothesis $H_0: r_{s,\alpha'}(z_i,M_i) =  0$
      based on bird data from 
      Bennett and Harvey~\citep{bennett87}
      (see  Figure~\ref{fig:truemet.heusner_resid_test} for details).
      Here, the individual plots correspond
      to the following ranges: (a) $M<0.1$ kg, (b) $M<1$ kg, (c) $M<10$ kg, (d) all birds.
      As for the mammal data, $p_{2/3}>0.05$ and $p_{3/4} \ll 10^{-4}$ for all mass ranges considered.
      }
}
{
    \caption{Test of the null hypothesis $H_0: r_{s,\alpha'}(z_i,M_i) =  0$
      based on bird data from 
      \citet{bennett87}
      (see  Figure~\ref{fig:truemet.heusner_resid_test} for details).
      Here, the individual plots correspond
      to the following ranges: (a) $M<0.1$ kg, (b) $M<1$ kg, (c) $M<10$ kg, (d) all birds.
      As for the mammal data, $p_{2/3}>0.05$ and $p_{3/4} \ll 10^{-4}$ for all mass ranges considered.
      }
}
    \label{fig:truemet.bennett_resid_test}
  \end{center}
\end{figure}

In this case we have tested the hypothesis for a
range of exponents, $\alpha'=0.6$--$0.8$, and calculated
the significance levels for both mammal and bird data
compiled by 
\ifthenelse{\boolean{revtexswitch}}{Heusner~\citep{heusner91}}{\citet{heusner91}}
and \ifthenelse{\boolean{revtexswitch}}{Bennett and Harvey~\citep{bennett87}}{\citet{bennett87}},
over different mass ranges.  
The results of this hypothesis test for Heusner's data is
contained in Figure~\ref{fig:truemet.heusner_resid_test}
and for Bennett and Harvey's data in
Figure~\ref{fig:truemet.bennett_resid_test}.
Both tests show that the hypothesis $\alpha=3/4$ is
rejected while that of $\alpha=2/3$ is not rejected
over all mass ranges considered for both birds and mammals.  
This does not mean that $\alpha=2/3$ is the ``real'' exponent,
but rather that it, unlike $\alpha=3/4$, is not incompatible with the data.

\section{Theories}
Thus far we have presented empirical evidence 
that $\alpha$ is mass dependent
and that the null hypothesis $\alpha=2/3$ should not be rejected for
mammals with $M < 10$ kg and all birds in most available data sets.
What then of theoretical attempts to derive
the 3/4-law of metabolism?
We show below that many of these arguments, while often elegant in conception and 
based on simple physics and geometry,
contain sufficient flaws to
render them unconvincing for the rejection of the
simplest theoretical hypothesis, $\alpha=2/3$.

\subsection{Dimensional analysis}
Dimensional analysis is a very useful technique
when there is only one mass, length, and time scale 
in a given problem.  However in the case of metabolic
scaling in biological
organisms there has been a long history of theoretical
debates over which scales to use when predicting
the scaling of metabolic rate via dimensional analysis.

Theories of biological and elastic similarities have been
used to explain many structural aspects of
organisms such as the length and width of major 
limbs~\citep{gunther82,economos82,gunther85}.
Using the principles of elastic similarity,
\ifthenelse{\boolean{revtexswitch}}{Bonner and McMahon~\citep{bonner83}}{\citet{bonner83}}
have tried to explain why
quarter-power scaling in body lengths and widths should lead to
$\alpha=3/4$.  
Cross-sections of limbs are argued to scale
as $M^{3/4}$ and therefore the power required to move
scales in the same way.
However, it is not clear why
the power output of muscles should be the dominant
factor in the scaling of basal metabolic rate.
Furthermore, such quarter-power scaling for animal shape
is not generally observed~\citep{calder84}.

Recent debates have focused on 
deriving $\alpha$ solely from dimensional analysis~\citep{heusner82a,feldman95}.  
The problem with all attempts to derive metabolic rate
from dimensional analysis is that different constraints
lead to different choices of contributing scales~\citep{feldman95}.
Explaining the scaling of metabolic rate is therefore displaced
to biological questions of energetic constraints, mass density,
physiological time, and diffusion constants across surfaces.

\subsection{Nutrient Supply Networks}
Interest in Kleiber's law resurged with the 
suggestion by \ifthenelse{\boolean{revtexswitch}}{West, Brown and Enquist (WBE)~\citep{west97}}{West, Brown and Enquist (WBE) (\citeyear{west97})}
that nutrient-supplying networks 
might be the ubiquitous limiting factor in
organismal form.
This remains an appealing and elegant idea
and stands as one of the boldest and most significant attempts at
discerning the underlying
physical mechanisms responsible for quarter-power scaling.
Although previous work had addressed the problem of optimal network
structure~\citep{cohn54a,cohn54b,rashevsky62,mlabarbera90},
theoretical relations between optimal networks and the
scaling of basal metabolic rate had never been considered.

The basic assumptions of WBE are 
\emph{i}) homoiotherms have evolved to 
minimize the rate at which they dissipate energy;
\emph{ii}) the relevant energy dissipation arises from transport
through nutrient-supply networks;
\emph{iii}) these networks are space-filling; 
and 
\emph{iv}) all homoiotherms possess
capillaries invariant in size.
From these four assumptions WBE derive three
important conclusions: 
\emph{i}) nutrient-supply networks are fractal;
\emph{ii}) these networks contain area-preserving branching; 
and
\emph{iii}) metabolic rate scales with $\alpha=3/4$.
However, as we show below and in 
Appendices~\ref{sec:truemet.murray} 
and~\ref{sec:truemet.areapreserving},
the arguments used are mathematically
incorrect and as a consequence none of the 
above conclusions may be derived from
the explicit assumptions.  
Nevertheless, we find the
model appealing and potentially useful in understanding
a number of biological issues.
Thus we detail below where the errors lie 
to illuminate the path of future work.
For clarity, we use the same notation as WBE.
For each level $k$ in the network hierarchy
one has $N_k$ vessels each with length $l_k$ and radius $r_k$ with
$k=1$ being the aorta and $k=N$ being the capillary level.
Related important quantities are $n_k$, $\gamma_k$ and
$\beta_k$, the ratios of number, length and radius from
levels $k-1$ to $k$.

Central to the theory is the connection of these
network ratios to metabolic rate.
WBE find that $n_k=n$, $\beta_k=\beta$ and $\gamma_k=\gamma$
are all constants independent of $k$ and that
\begin{equation}
  \label{eq:truemet.alpha}
  \alpha = -\frac{\log n}{\log \gamma \beta^2}.
\end{equation}
This depends in part on an assumption, which we discuss below,
that $B \propto N_c$ where $N_c$ is the number of capillaries.
They also conclude that 
\begin{equation}
  \label{eq:truemetbetagamma}
  \beta=n^{-1/2}
  \qquad
  \mbox{and}
  \qquad
  \gamma=n^{-1/3},
\end{equation}
which gives $\alpha=3/4$ in equation~\req{eq:truemet.alpha}.
Whereas we show below
that these relations do not arise from
an optimization principle, they do
have simple interpretations.
The first relation corresponds
to networks being area-preserving 
via $N_k r_k^2 = N_{k-1} r_{k-1}^2$.
The second relation follows
from a space-filling criterion that $N_k l_k^3 = N_{k-1} l_{k-1}^3$.
Whether or not space-filling networks satisfy these
conditions has been discussed by
\ifthenelse{\boolean{revtexswitch}}{Turcotte \etal~\citep{turcotte98},}{\citet{turcotte98},}
who consider the more general case of
side-branching networks and arrive at an equivalent statement of equation~\req{eq:truemet.alpha}
where the network ratios $\beta$ and $\gamma$ are to be 
determined empirically as functions of $n$.

WBE minimize energy dissipation rate by
minimizing network impedance using a 
Lagrange multiplier method.
Two types of impedance are considered: Poiseuille flow~\citep{lamb45}
and, for the case of mammals and birds, 
a more realistic pulsatile flow~\citep{womersley55a}.

We use the Poiseuille case to demonstrate
how fractality is not proven by the minimization procedure.
The impedance is given by
\begin{equation}
Z = \sum_{k=0}^N \frac{8\mu l_k}{\pi r_k^4 N_k} = \sum_{k=0}^N Z_k
\label{eq:truemet.Zpoiseuille}
\end{equation}
where $Z_k$ is the effective impedance of the $k$th level.
As WBE show, the equations
are consistent and  $Z$ is minimized when
\begin{equation}
  \gamma_k = \beta_k = n_k^{-1/3}.
  \label{eq:truemet.poisratios}
\end{equation}
However, the calculations do not require
these ratios to be level-independent, and as a consequence,
the network need not be fractal.
Further details may be
found in Appendix~\ref{sec:truemet.murray}.
To see why this is true, we observe that
equations~\req{eq:truemet.Zpoiseuille} and~\req{eq:truemet.poisratios} give
\begin{equation}
  \label{eq:Z_kZ}
Z_k = \gamma_k\beta_k^{-4}n_k^{-1} Z_{k-1} = 1 \cdot Z_{k-1}.  
\end{equation}
Thus, $Z$, the quantity being minimized, is invariant
as long as $\gamma_k\beta_k^{-4}n_k^{-1} = 1$ for each $k$. 
This shows that in this setting, a network can have
$n_k$ varying with $k$ and still be ``efficient.''
A finding of fractal networks would have provided a derivation
of Murray's empirical law which essentially 
states that $\beta = n^{-1/3}$ for the outer reaches
of the cardiovascular system~\citep{murray26}.

Regardless of these issues, the assumption of Poiseuille flow
leads to an approximate metabolic scaling law with $\alpha=1$.
WBE suggest that modeling pulsatile
flow will provide the explanation for $\alpha=3/4$.
The impedance now takes the form
\begin{equation}
Z \propto \sum_{k=0}^N \frac{h_k^{1/2}}{\sqrt{2}\pi r_k^{5/2} N_k},
\label{eq:truemet.Zpulsatile}
\end{equation}
where $h_k$ is the thickness of the vessel wall.
However, as explained in Appendix~\ref{sec:truemet.areapreserving},
the equations given by the Lagrange
multiplier technique are inconsistent.
For example, the equations give $h_k = -r_k/5$ which means
negative wall thicknesses
for blood vessels when they are by definition
positive~\citep{womersley55a,womersley55b}.
If reasonable modifications are made to circumvent this issue,
then the equations lead to $\alpha=6/7$ rather than $\alpha=3/4$.

In order to obtain the scaling $\alpha=3/4$ one could 
abandon the minimization calculation and 
assume a fractal, space-filling, area-preserving network
where $B \propto N_c$.
In support of such an assumption, there is good empirical evidence
that blood systems are well
approximated by fractals~\citep{zamir83,fung90,kassab93a,kassab93b}.
With regards to the assumption that $B \propto N_c$, 
direct measurements for capillary density ($N_c/M \propto M^{\alpha-1}$) are 
reported by \ifthenelse{\boolean{revtexswitch}}{Hoppeler \etal~\citep{hoppeler81}}{\citet{hoppeler81}}
with exponents for
the scaling of capillary density across species 
ranging from $-0.21 \pm 0.04$ 
to $-0.07 \pm 0.11$ for various regions of muscle.
These numbers are in keeping with
higher exponents for the scaling of $N_c$ with $M$
in the range 0.75--1.00, but whether or not $B\propto N_c$ is
itself an unproven assumption.  It is probably more
likely that the number of capillaries scales with the maximum
metabolic rate which is thought to scale with an exponent
closer to unity~\citep{bishop99}.  
At rest not all capillaries diffuse oxygen
simultaneously and the limiting factor
for basal metabolic rate might not be $N_c$.

A simpler and more recent theory based on the idea of networks
has been proposed by 
\ifthenelse{\boolean{revtexswitch}}{Banavar \etal~\citep{banavar99}.}{\citet{banavar99}.}
Here, networks fill $D$-dimensional hypercubes that
have $L^D$ uniformly distributed transfer sites.
The theory is applied to both three-dimensional organisms
and two-dimensional river networks.
For organisms, Banavar \etal\ find blood volume scales
as $V_b \propto L^{(D+1)}$.  
Since Banavar \etal\ further assume that $B \propto L^D$
and that $V_b \propto M$, they conclude
that $B \propto M^{D/(D+1)}$.  Thus, when $D=3$, this gives
$\alpha=3/4$.

However, transfer sites are assumed to be invariant in size
and hence $L^D$ appears to be proportional to volume $V$
consequently $M$.  
Thus, both the scalings $V_b \propto M$
and $V_b \propto M^{(D+1)/D}$ are used, creating an
apparent inconsistency.   The scaling of the distance between
transfer sites and the distinction between Euclidean
and non-Euclidean length scales could possibly be 
clarified to help resolve the
dilemma.  Note that $V_b \propto M$ is
supported empirically~\citep{stahl67}.
\subsection{Four-dimensional biology}
Over two decades ago it was suggested by \ifthenelse{\boolean{revtexswitch}}{Blum~\citep{blum77}}{\citet{blum77}}
that $\alpha=3/4$ could
be understood by appealing to a surface law of metabolism in
a four-dimensional space.  In $d$ dimensions, the ``area'' $A$ of the
hypersurface enclosing a $d$-dimensional hypervolume scales like
$A\propto V^{\frac{d-1}{d}}$.  When $d=4$, $A\propto V^{3/4}$, although
how this could be reconciled with our three-dimensional world was
not explained and the theory has been refuted elsewhere~\citep{speakman90}.

Recently, an attempt
by \ifthenelse{\boolean{revtexswitch}}{West \etal~\citep{west99}}{\citet{west99}}
has been made to refine and generalize their earlier work on metabolic
scaling~\citep{west97}
using an optimization procedure to
explain how an effective fourth dimension could yield $\alpha=3/4$.
The idea put forward is that organisms have evolved to maximize the scaling
of the effective surface area, $a$,
across which resources are exchanged.
The area $a$ and the biological volume $v$ are shown 
to satisfy the relation
\begin{equation}
  \label{eq:truemet.avrel}
  a \propto v^{(2+\epsilon_a)/(3+\epsilon_v)},
\end{equation}
where $\epsilon_a$ and $\epsilon_v$ are
exponents to be determined by optimization.
West \etal\ then introduce the relationship
$v = al$ where $l$ is a characteristic length 
of the organism.  With the further assumption
that $v \propto M$, equation~\req{eq:truemet.avrel}
then becomes
\begin{equation}
  \label{eq:truemet.aMrel}
  a \propto M^{(2+\epsilon_a)/(3+\epsilon_a+\epsilon_l)},
\end{equation}
where $\epsilon_l = \epsilon_v - \epsilon_a$. 
With the conditions that $0 \leq \epsilon_l, \epsilon_a \leq 1$,
West \etal\ find that $\epsilon_a = 1$ and $\epsilon_l = 0$.
Equation~\req{eq:truemet.aMrel} then yields $a \propto M^{3/4}$.
Assuming $a \propto B$, this gives $\alpha=3/4$.

However, this result contradicts
the geometric fact that transfer area can maximally
scale as volume, i.e., $a \propto v$, which gives $\alpha=1$.
Indeed, this result is obtained by 
optimizing equation~\req{eq:truemet.avrel}
instead of equation~\req{eq:truemet.aMrel}.
Doing so leads to $\epsilon_a = 1$ and $\epsilon_v = 0$,
assuming $0 \leq \epsilon_a, \epsilon_v \leq 1$,
which gives $a \propto M$, i.e., $\alpha=1$.
In order to reconcile this
with the results of West \etal, we note that the
bounds $0 \leq \epsilon_l, \epsilon_a, \epsilon_v \leq 1$
are overly restrictive.  
For example, $\epsilon_l = -1$
corresponds to the relevant length $l$ being invariant with
respect to $M$ and,
in this case, equation~\req{eq:truemet.aMrel}
then gives the same scaling as~\req{eq:truemet.avrel},
namely, $a \propto M$.
Thus, the contradiction is resolved
and the optimization procedure is seen to yield $\alpha=1$
rather than $\alpha=3/4$.

\section{Discussion}

The possibility that there might be a simple
law to explain the scaling of metabolic rates still captures the
imagination of many seeking to understand 
what Kleiber called ``the fire of life''~\citep{kleiber61}.
It is perhaps for this reason that so many researchers,
theorists and empiricists alike, have struggled to deduce
explanations for the deviations from the simplest
expectation that $\alpha=2/3$.  

The shift from $\alpha=2/3$ to $\alpha=3/4$
began with the early work by Kleiber and Brody who found
$\alpha\simeq$ 0.72--0.73 in limited data sets~\citep{kleiber32,brody45}.
Afterwards it was work by \ifthenelse{\boolean{revtexswitch}}{Hemmingsen~\citep{hemmingsen60}}{\citet{hemmingsen60}}
and a general consensus among 
practitioners~\citep{blaxter65} that
simple fractions would be a more convenient standard that
led to the widespread acceptance of $\alpha=3/4$.  
Subsequently, $\alpha=3/4$ has often been taken as fact despite the
absence of a comprehensive theory and
contradictory evidence from large literature surveys.  
Most prominent among these surveys are those 
by \ifthenelse{\boolean{revtexswitch}}{Bartels~\citep{bartels82},}{\citet{bartels82},} \ifthenelse{\boolean{revtexswitch}}{Bennett and Harvey~\citep{bennett87},}{\citet{bennett87},}
and \ifthenelse{\boolean{revtexswitch}}{Heusner~\citep{heusner91},}{\citet{heusner91},}
which suggest that $\alpha$ depends on body size and taxonomic
level.

We have re-analyzed some of the most influential
empirical data sets in the study of metabolic rate
scaling.  We have constructed a set of hypothesis
tests which show that in the data sets
of Kleiber, Brody, Bennett and Harvey, 
and Heusner, pure $3/4$-law scaling is not present.
For both mammals with $M \leq 10$ kg and all birds
we are unable to reject the null hypothesis $\alpha=2/3$.
For mammals with $M \geq 10$ kg, systematic deviations from $\alpha=2/3$ appear
to be present in all of the data sets, the roots of which
might simply be a consequence of a change in body shape for large mammals
or might point to a greater evolutionary advantage of 
large mammals.  

We have also reviewed historic and recent attempts
to justify $\alpha=3/4$ theoretically. Many of the
early efforts to explain the scaling of metabolic  
rates via dimensional analysis and other crude scaling
techniques have been dismissed in the past.  
Although recent attempts to link metabolic rates to network
structure are noteworthy they do not prove the
stated conclusions.  Nonetheless, we believe that
research exploring the role of geometric form and
the dynamics of growth in constraining the behavior of 
networks might lead to important progress in 
organismal biology.

Stated simply, after a systematic
review of the available empirical data and theoretical
arguments, we find no compelling evidence of 
a simple scaling law for metabolic rate, and if it were to
exist, we also find no compelling evidence that the exponent
should be $\alpha=3/4$.

\section*{Acknowledgements}
We would like to offer our sincere thanks to A.~Heusner for 
helpful discussions and sharing
data with us.  We would also like
to thank M.~Brenner, A.~Brockwell, L.~Demetrius, H.~Fraser,
H.~Hartman,  K.~Schmidt-Nielsen,
and N.~Schorghofer for their comments.
PSD and DHR thank G.~West for hosting visits to 
Los Alamos National Laboratory and the Santa Fe Institute,
thereby aiding our introduction to the subject.
We would also like to thank M.~Kardar, A.~Rinaldo,
and the other participants in the 1999 MIT seminar
on natural networks for their insightful discussions.
This work was supported in part by NSF grant EAR-9706220
and DOE grant DEF602-99ER15004.  JSW is grateful
for support from an NDSEG fellowship.

\appendix

\section{Network optimization calculation for Poiseuille flow}
\label{sec:truemet.murray}

We follow the conventions of WBE
and consider the case of Poiseuille flow as a means
to derive Murray's law and the
fractal nature of nutrient supply networks.  
The impedance of the network is
\begin{equation}
Z = \sum_{k=0}^N \frac{8\mu l_k}{\pi r_k^4 N_k}.
\label{eq:truemet.Zpoiseuille2}
\end{equation}
Minimizing the network's impedance with the
Lagrange constraints of fixed mass and blood volume
along with the assumption of a space filling network
leads to the auxiliary function,
\ifthenelse{\boolean{twocolswitch}}
{
  \begin{eqnarray}
    \lefteqn{F_m(r_k,l_k,N_k,M) = \sum_{k=0}^N \frac{8\mu l_k}{\pi r_k^4 N_k}} \nonumber \\ 
    & + & \lambda \sum_{k=0}^{N} \pi r_k^2 l_k N_k
    + \sum_{k=0}^{N} \lambda_k N_k l_k^3
    + \lambda_M M.
    \label{eq:truemet.lagrangefn-murray}
  \end{eqnarray}
  }
{
  \begin{equation}
    F_m(r_k,l_k,N_k,M)
  = \sum_{k=0}^N \frac{8\mu l_k}{\pi r_k^4 N_k}
  + \lambda \sum_{k=0}^{N} \pi r_k^2 l_k N_k
  + \sum_{k=0}^{N} \lambda_k N_k l_k^3
  + \lambda_M M.
  \label{eq:truemet.lagrangefn-murray}
\end{equation}
}
Taking partial derivatives with respect to $l_j$, $r_j$ and $N_j$
we respectively have
\begin{eqnarray}
\partialdiff{F_m}{l_j} & = &
\frac{8\mu}{\pi r_j^4 N_j}
+ \lambda \pi r_j^2 N_j
+3\lambda_j N_j l_j^{\, 2}
= 0, 
\label{eq:truemet.murray_l}\\
\partialdiff{F_m}{r_j} & = &
\frac{-4\cdot 8\mu l_j}{\pi r_j^5 N_j}
+ \lambda 2\pi r_j l_j N_j 
= 0, 
\label{eq:truemet.murray_r}
\end{eqnarray}
and
\begin{eqnarray}
\partialdiff{F_m}{N_j} & = &
\frac{-1\cdot 8\mu l_j}{\pi r_j^4 N_j^2}
+ \lambda \pi r_j^2 l_j
+ \lambda_j l_j^{\, 3}
= 0.
\label{eq:truemet.murray_N}
\end{eqnarray}
Considering first equation~\req{eq:truemet.murray_r}, 
we obtain 
\begin{equation}
\lambda = \frac{16\mu}{\pi^2 r_j^6 N_j^2}.
\label{eq:truemet.murray_r_manip1}
\end{equation}
Since this holds for all $j$ then
\begin{equation}
1 = \frac{16 \mu}{\pi^2 r_{j-1}^6 N_{j-1}^2} \frac{\pi^2 r_{j}^6 N_j^2}{16 \mu}
  = \beta_j^{\, 6} n_j^2,
\label{eq:truemet.murray_beta_n_manip}
\end{equation}
where $\beta_j=r_j/r_{j-1}$ and $n_j=N_j/N_{j-1}$
which demonstrates that
\begin{equation}
\beta_j = n_j^{-1/3},
\label{eq:truemet.murray_beta_n}
\end{equation}
giving us Murray's law~\citep{murray26}.

After rearranging 
equation~\req{eq:truemet.murray_l} we obtain
\begin{align}
\lambda_j  & = -\frac{8\mu}{3 \pi r_j^4 N_j^2 l_j^{\, 2}} 
 - \frac{\lambda \pi r_j^2}{3 l_j^{\, 2}} \nonumber \\
 & = -\frac{8\mu}{3 \pi r_j^4 N_j^2 l_j^{\, 2}}
-\frac{16\mu}{3 \pi r_j^4 N_j^2 l_j^{\, 2}} \nonumber \\
 & = -\frac{8\mu}{ \pi r_j^4 N_j^2 l_j^{\, 2}}, 
\label{eq:truemet.murrayl_manip}
\end{align}
where we have used the form for $\lambda$ obtained
in equation~\req{eq:truemet.murray_r_manip1}.  
Note that derivatives with respect to $N_j$, equation~\req{eq:truemet.murray_N}, 
yield the same expression for $\lambda_j$ given above:
\begin{align}
\lambda_j & = 
-\frac{-1\cdot 8\mu}{\pi r_j^4 N_j^2 l_j^{\, 2}}
-\frac{\lambda \pi r_j^2}{l_j^{\, 2}} \nonumber \\
 & =\frac{8\mu}{\pi r_j^4 N_j^2 l_j^{\, 2}}
-\frac{16\mu}{\pi r_j^4 N_j^2 l_j^{\, 2}} \nonumber \\
 & = - \frac{8\mu}{\pi r_j^4 N_j^2 l_j^{\, 2}}
\label{eq:truemet.murrayN_manip}
\end{align}
The three equations~\req{eq:truemet.murray_l}, \req{eq:truemet.murray_r} and~\req{eq:truemet.murray_N}
are therefore consistent but redundant.  The redundancy
can be seen to lie in the fact that the auxiliary function $F_m$
in equation~\req{eq:truemet.lagrangefn-murray} can be written
in terms of only two variables for each level $k$: 
$\xi_k = N_k l_k^{\, 3}$ and $\zeta_k = r_k/l_k$.
Equation~\req{eq:truemet.lagrangefn-murray}
thus becomes
\ifthenelse{\boolean{twocolswitch}}
{
  \begin{eqnarray}
    \lefteqn{F_m(\xi_k,\zeta_k,M) = \sum_{k=0}^N \frac{8\mu}{\pi \zeta_k^4 \xi_k}} \nonumber \\
    & + & \lambda \sum_{k=0}^{N} \pi \zeta_k^2 \xi_k
    + \sum_{k=0}^{N} \lambda_k \xi_k
    + \lambda_M M.
    \label{eq:truemet.lagrangefn-murray-2vars}
  \end{eqnarray}
  }
{
  \begin{equation}
    F_m(\xi_k,\zeta_k,M)
    = \sum_{k=0}^N \frac{8\mu}{\pi \zeta_k^4 \xi_k}
    + \lambda \sum_{k=0}^{N} \pi \zeta_k^2 \xi_k
    + \sum_{k=0}^{N} \lambda_k \xi_k
    + \lambda_M M.
    \label{eq:truemet.lagrangefn-murray-2vars}
  \end{equation}
  }
Thus one is only able to obtain information such
as ratios of variables rather than exact values for
network parameters.

The scaling of length ratios are explicitly determined by
WBE's space-filling assumption
\begin{equation}
N_k l_k^{\, 3} = C.
\label{eq:truemet.spacefilling}
\end{equation}
Thus, even without implementing the minimization procedure the
space-filling assumption implies
\begin{equation}
\gamma_k = n_k^{-1/3},
\label{eq:truemet.murray-gamma-n}
\end{equation}
where $\gamma_k = l_k/l_{k-1}$ is the length ratio.
Finally, the equations~\req{eq:truemet.murray_beta_n},
\req{eq:truemet.murrayl_manip} (or ~\req{eq:truemet.murrayN_manip})
and~\req{eq:truemet.murray-gamma-n} combine to give
\ifthenelse{\boolean{twocolswitch}}
{
\begin{eqnarray}
\frac{\lambda_k}{\lambda_{k-1}}
& = & \frac{r_{k-1}^4 N_{k-1}^2 l_{k-1}^{\, 2}}{r_k^4 N_k^2 l_k^{\, 2}} \nonumber \\
& = & \beta_k^{-4} n_k^{-2} \gamma_k^{-2} \nonumber \\
& = & (n_k^{-1/3})^{-4} n_k^{-2} (n_k^{-1/3})^{-2} = 1
\label{eq:truemet.}
\end{eqnarray}
}
{
\begin{equation}
\frac{\lambda_k}{\lambda_{k-1}}
= 
\frac{r_{k-1}^4 N_{k-1}^2 l_{k-1}^{\, 2}}{r_k^4 N_k^2 l_k^{\, 2}} 
= \beta_k^{-4} n_k^{-2} \gamma_k^{-2}
= (n_k^{-1/3})^{-4} n_k^{-2} (n_k^{-1/3})^{-2} = 1
\label{eq:truemet.}
\end{equation}
}
so we have $\lambda_k = \lambda_0$ for all $k$.

The calculations are seen to be consistent and yield Murray's
law~\citep{murray26}. 
Variations with respect to $M$ are more subtle
since $N=N(M)$ and provide higher order corrections.
However, one of WBE's crucial results,
$n_k = n$, i.e.\ that the network is fractal,
has not been reproduced.  
One way to see this is to consider the impedance as impedances in series:
\begin{equation}
Z = \sum_{k=0}^N Z_k \ \
\mbox{where}
\ \ Z_k =  \frac{8\mu l_k}{\pi r_k^4 N_k}.
\label{eq:truemet.Zsum}
\end{equation}
Using equations~\req{eq:truemet.murray_beta_n} 
and~\req{eq:truemet.murray-gamma-n} we have that
\begin{equation}
\frac{Z_k}{Z_{k-1}}
= \frac{r_{k-1}^4 N_{k-1} l_k}{r_{k}^4 N_{k} l_{k-1}}
= \frac{\gamma_k}{\beta_k^4 n_k}
= \frac{n_k^{-1/3}}{(n_k^{-1/3})^4 n_k} = 1.
\label{eq:truemet.Zratio}
\end{equation}
In other words, the same impedance appears at each level.
So 
\begin{equation}
Z = (N+1)Z_N \simeq N Z_N = N \frac{8\mu l_c}{\pi r_c^4 N_c}\propto \frac{N}{N_c},
\label{eq:truemet.Zinvariance}
\end{equation}
since $r_c$ and $l_c$ are assumed to be independent of mass
and $N_c$ is the number of capillaries.
This is true regardless of whether or not the 
structure is fractal.  The network has to possess
branching ratios that collectively maintain
the same impedance from level to level 
(i.e., $\gamma_k/ \beta_k^4 n_k = 1$ as per 
equation~\req{eq:truemet.Zratio})
but there is no requirement that the individual
ratios $\gamma_k$, $\beta_k$ and $n_k$ be independent of level.
Moreover, without the result that the network is fractal, this
minimization procedure
no longer yields the power law scaling of metabolic
rate (see equation~\req{eq:truemet.alpha}).

\section{Network optimization calculation for pulsatile flow}
\label{sec:truemet.areapreserving}

In the case of Poiseuille flow, WBE
find a network structure where
area preservation is not satisfied ($\beta_k \neq n_k^{-1/2}$) and,
effectively, $\alpha=1$ (if $n_k = n$ is assumed).  
The intended fix is to properly 
treat pulsatile flow of mammalian
blood circulation systems.  
By doing so we should obtain 
$\beta_k = n_k^{-1/2}$ and $\gamma_k  = n_k^{-1/3}$.
Together with the assumption $n_k = n$, this leads
to the conclusion, $N_c\propto M^{3/4}$, and assuming
$B\propto N_c$, it would imply a $3/4$-law of metabolic scaling.

The calculation relies on the results of Womersley's work on
pulsatile flow~\citep{womersley55a,womersley55b}.  
Womersley's calculations lead to a modification
of the Poiseuille impedance.  For large tubes one has
\begin{equation}
Z \simeq \frac{\rho c_0}{\pi r^2}.
\label{eq:truemet.Zpulsatile2}
\end{equation}
where $c_o = (Eh/2\rho r)^{1/2}$ is the Korteweg-Moens velocity,
$E$ is Young's modulus, $h$ is the thickness of the vessel wall, 
$\rho$ is blood density and $r$ is, as before, the 
vessel radius~\citep{womersley55a,womersley55b}.  
This impedance appears to be per unit length
but it has the correct dimensions showing that for a flow
with pulsatile forcing in an elastic tube, the impedance is
independent of the tube length.

Womersley's impedance suggests a new auxiliary function:
\ifthenelse{\boolean{twocolswitch}}
{
  \begin{multline}
    F_w(r_k,h_k,l_k,N_k,M) = \sum_{k=0}^N \frac{(E h_k \rho)^{1/2}}{\sqrt{2}\pi r_k^{5/2} N_k} \\
    + \lambda \sum_{k=0}^{N} \pi (r_k+h_k)^2 l_k N_k
    + \sum_{k=0}^{N} \lambda_k N_k l_k^3
    + \lambda_M M.
    \label{eq:truemet.lagrangefn-womersley}
  \end{multline}
  }
{
  \begin{multline}
    F_w(r_k,h_k,l_k,N_k,M) \\
    = \sum_{k=0}^N \frac{(E h_k \rho)^{1/2}}{\sqrt{2}\pi r_k^{5/2} N_k}
    + \lambda \sum_{k=0}^{N} \pi (r_k+h_k)^2 l_k N_k
    + \sum_{k=0}^{N} \lambda_k N_k l_k^3
    + \lambda_M M.
    \label{eq:truemet.lagrangefn-womersley}
  \end{multline}
  }
Note that the extra variable of wall thickness, $h_k$, has been included
in the second term to make it a measure
of the total volume taken up by the blood system.
The variable $h_k$ must appear in the constraints
if the minimization is to make 
any sense and the blood volume is
the only reasonable choice---the blood volume constraint becomes
a network volume constraint.

On considering variations of equation~\req{eq:truemet.lagrangefn-womersley}
with respect to $r_j$ and $h_j$ we obtain
\begin{equation}
\partialdiff{F_w}{r_j} =
-\frac{5}{2}\frac{(E h_j \rho)^{1/2}}{\sqrt{2}\pi r_j^{7/2} N_j}
+ \lambda 2 \pi (r_j + h_j) l_j N_j
= 0,
\label{eq:truemet.womersley_r}
\end{equation}
and
\begin{equation}
\partialdiff{F_w}{h_j} =
\frac{1}{2}\frac{(E\rho)^{1/2}}{\sqrt{2}\pi h_j^{1/2}r_j^{5/2} N_j}
+ \lambda 2 \pi (r_j + h_j) l_j N_j
= 0.
\label{eq:truemet.womersley_h}
\end{equation}
Since the second term of these equations are the same we
then have an equality between the first terms which simplifies to show that
\begin{equation}
h_j = -\frac{1}{5} r_j.
\label{eq:truemet.womersley_r_h_manip}
\end{equation}

This suggests that $r_k$ is the distance to the
outer wall of blood vessels.  We are then measuring
the blood volume as before and we should have had $(r_k-h_k)$ instead
of $(r_k+h_k)$ in the auxiliary function.
\ifthenelse{\boolean{revtexswitch}}{However it is apparent from Womersley's 
work~\citep{womersley55a,womersley55b}}{However it is apparent from \citet{womersley55a,womersley55b}}
that $r$ is the radius as measured from the center to the inner wall
of a blood vessel rather than the outer wall.  
There appears to be
no reasonable and simple way of
including the $h_k$ into a constraint function and we 
have an ill-posed problem.

Nevertheless, we may proceed with the calculation by 
adding an extra assumption that
$h_k = a_0 r_k$ where $a_0 > 0$.
The modified version of 
equation~\req{eq:truemet.womersley_r} 
now gives $\lambda$ as
\begin{equation}
\lambda = \frac{(a_0 E \rho)^{1/2}}{\sqrt{2}\pi^2 (1+a_0)^2 r_j^4 N_j^2 l_j}.
\label{eq:truemet.womersley_lambda}
\end{equation}
Since the right hand side is independent of $j$ we must
therefore have
\begin{equation}
\beta_j^4 n_j^2 \gamma_j = 1,
\label{eq:truemet.womersley_ratios1}
\end{equation}
and given the space-filling constraint,
$\gamma_j = n_j^{-1/3}$, we obtain
\begin{equation}
1 =\beta_j^4 n_j^2 n_j^{-1/3} = \beta_j^4 n_j^{5/3},
\label{eq:truemet.womersley_ratios2}
\end{equation}
which gives a relationship between the radius and number
ratios that is not area-preserving:
\begin{equation}
\beta_j = n_j^{-5/12}.
\label{eq:truemet.womersley_ratios3}
\end{equation}
A further complication here is that the equations obtained
by setting $\tpartialdiff{F_w}{l_j}=0$
and $\tpartialdiff{F_w}{N_j}=0$ are not consistent.

As in the case of Poiseuille flow, $n_k=n$ is not derivable.
Assuming that $n_k = n$ and using equation~\req{eq:truemet.alpha}
we find that the metabolic exponent should be
\begin{equation}
\alpha = -\frac{\ln n}{\ln \gamma \beta^2}
= -\frac{\ln n}{\ln n^{-1/3} n^{-10/12}}
= 6/7,
\label{eq:truemet.6/7law}
\end{equation}
as opposed to the stated $3/4$.

Note that if we had found $\beta_k=n_k^{1/2}$ then
the $3/4$ law would have been deduced (again, assuming $n_k=n$).  
Another observation
here is that if the Womersley impedance is taken together
with $\beta_k=n_k^{-1/2}$ then we find that
the minimum total impedance is obtained irrespective
of the ratios $n_k$ being equal or not.  So, in the cases
of Poiseuille and pulsatile flow a fractal network
is not necessary for energy dissipation to be minimized.
Additionally, in the case of a pulsatile flow network,
$\alpha=3/4$ cannot be derived from the optimization problem
as stated.  It may instead be derived by assuming an area preserving,
space-filling, fractal network where $B\propto N_c$.


\begin{thebibliography}{10}

\bibitem{bonner83}
J.~T. Bonner and T.~A. McMahon.
\newblock {\em On Size and Life}.
\newblock Scientific American Library, New York, 1983.

\bibitem{calder84}
W.~A. Calder~III.
\newblock {\em Size, Function and Life History}.
\newblock Dover, New York, 1996.

\bibitem{kleiber32}
M.~Kleiber.
\newblock Body size and metabolism.
\newblock {\em Hilgardia}, 6:315--353, 1932.

\bibitem{kleiber61}
M.~Kleiber.
\newblock {\em The Fire of Life. An Introduction to Animal Energetics}.
\newblock Wiley, New York, 1961.

\bibitem{peters83}
R.~Peters.
\newblock {\em The Ecological Implications of Body Size}.
\newblock Cambridge University Press, Cambridge, 1983.

\bibitem{schmidt-nielsen84}
K.~Schmidt-Nielsen.
\newblock {\em Scaling: Why is Animal Size So Important?}
\newblock Cambridge University Press, UK, 1984.

\bibitem{anderson97}
B.~J. Anderson, A.~D. Mc{K}ee, and N.~H.~G. Holford.
\newblock Size, myths, and the clinical pharmacokinetics of analgesia in
  paediatric patients.
\newblock {\em Clinical Pharmacokinetics}, 33:313--27, 1997.

\bibitem{mordenti86}
J.~Mordenti.
\newblock Man versus beast: pharmacokinetic scaling in mammals.
\newblock {\em Journal of Pharmaceutical Sciences}, 75:1028--1040, 1986.

\bibitem{mahmood99}
I.~Mahmood.
\newblock Allometric issues in drug development.
\newblock {\em Journal of Pharmaceutical Sciences}, 88:1101--6, 1999.

\bibitem{burger91}
I.~Burger and J.~Johnson.
\newblock Dogs large and small: the allometry of energy requirements within a
  single species.
\newblock {\em J. Nutr.}, 121(11 (Suppl)):S18--21, 1991.

\bibitem{cunningham80}
J.~Cunningham.
\newblock A reanalysis of the factors influencing basal metabolic rate in
  normal adults.
\newblock {\em Am. J. Clin. Nutr.}, 33:2372--2374, 1980.

\bibitem{pike84}
R.~Pike and M.~Brown.
\newblock {\em Nutrition an integrated approach}.
\newblock John Wiley and Sons, New York, 1984.

\bibitem{damuth81}
J.~Damuth.
\newblock Population density and body size in mammalas.
\newblock {\em Nature}, 290:699--700, April 1981.

\bibitem{lindstedt86}
S.~L. Lindstedt, B.~J. Miller, and S.~W. Buskirk.
\newblock Home range, time, and body size in mammals.
\newblock {\em Ecology}, 67(2):413--418, 1986.

\bibitem{carbone99}
C.~Carbone, G.~Mace, C.~Roberts, and D.~Macdonald.
\newblock Energetic constraints on the diet of terrestial carnivores.
\newblock {\em Nature}, 402:286--8, 1999.

\bibitem{economos83}
A.~E. Economos.
\newblock Elastic and/or geometric similarity in mammalian design.
\newblock {\em J. Theor. Biol.}, 103:167--172, 1983.

\bibitem{feldman82}
H.~Feldman.
\newblock The 3/4 mass exponent for energy metabolism is not a statistical
  artifact.
\newblock {\em Respir. Physiol.}, 52:149--163, 1983.

\bibitem{feldman95}
H.~A. Feldman.
\newblock On the allometric mass exponent, when it exists.
\newblock {\em J. Theor. Biol.}, 172:187--197, 1995.

\bibitem{heusner82b}
A.~Heusner.
\newblock Energy metabolism and body size. {I}.\ {I}s the 0.75 mass exponent of
  {K}leiber a statistical artifact?
\newblock {\em Respir. Physiol.}, 48:1--12, 1982.

\bibitem{prothero84a}
J.~Prothero.
\newblock Scaling of standard energy-metabolism in mammals: {I}. {N}eglect of
  circadian-rhythms.
\newblock {\em J. Theor. Biol.}, 106(1):1--8, 1984.

\bibitem{blaxter65}
K.~L. Blaxter, editor.
\newblock {\em Energy Metabolism; Proceedings of the 3rd symposium held at
  Troon, Scotland, May 1964}.
\newblock Academic Press, New York, 1965.

\bibitem{west97}
G.~B. West, J.~H. Brown, and B.~J. Enquist.
\newblock A general model for the origin of allometric scaling laws in biology.
\newblock {\em Science}, 276:122--126, April 1997.

\bibitem{hemmingsen60}
A.~Hemmingsen.
\newblock Energy metabolism as related to body size and respiratory surfaces,
  and its evolution.
\newblock {\em Rep. Steno Mem. Hosp.}, 9:1--110, 1960.

\bibitem{heusner87}
A.~Heusner.
\newblock What does the power function reveal about structure and function in
  animals of different size?
\newblock {\em Ann. Rev. Phsyiol.}, 49:121--133, 1987.

\bibitem{rubner1883}
M.~Rubner.
\newblock Ueber den einfluss der k\"{o}rpergr\"{o}sse auf stoffund
  kraftwechsel.
\newblock {\em Z. Biol.}, 19:535--562, 1883.

\bibitem{brody45}
S.~Brody.
\newblock {\em Bioenergetics and Growth}.
\newblock Reinhold, New York, 1945.

\bibitem{schmidt-nielsen84note}
Kleiber's motivation in part was to make calculations less cumbersome with a
  slide rule (see p.\ 59 in~\cite{schmidt-nielsen84}).

\bibitem{heusner91}
A.~A. Heusner.
\newblock Size and power in mammals.
\newblock {\em J. Exp. Biol.}, 160:25--54, 1991.

\bibitem{bennett87}
P.~Bennett and P.~Harvey.
\newblock Active and resting metabolism in birds---allometry, phylogeny and
  ecology.
\newblock {\em J. Zool.}, 213:327--363, 1987.

\bibitem{bartels82}
H.~Bartels.
\newblock Metabolic rate of mammals equals the 0.75 power of their body weight.
\newblock {\em Exp. Biol. and Medicine}, 7:1--11, 1982.

\bibitem{economos82}
A.~C. Economos.
\newblock On the origin of biological similarity.
\newblock {\em J. Theor. Biol.}, 94:25--60, 1982.

\bibitem{gunther85}
B.~Gunther.
\newblock Theories of biological similarities---30 years of trial and error.
\newblock {\em Arch. Biol. Med. Exp.}, 18(3--4):197--224, 1985.

\bibitem{gunther82}
B.~Gunther and E.~Morgado.
\newblock Dimensional analysis and theory of biological similarity.
\newblock {\em Exp. Biol. Med.}, 7:12--20, 1982.

\bibitem{heusner82a}
A.~Heusner.
\newblock Energy metabolism and body size. {I}{I}.\ {D}imensional analysis and
  energetic non-similarity.
\newblock {\em Respir. Physiol.}, 48:13--25, 1982.

\bibitem{blum77}
J.~J. Blum.
\newblock On the geometry of four-dimensions and the relationship between
  metabolism and body mass.
\newblock {\em J. Theor. Biol.}, 64:599--601, 1977.

\bibitem{west99}
G.~B. West, J.~H. Brown, and B.~J. Enquist.
\newblock The fourth dimension of life; fractal geometry and allometric scaling
  of organisms.
\newblock {\em Nature}, 284:1677--1679, 1999.

\bibitem{banavar99}
J.~R. Banavar, A.~Maritan, and A.~Rinaldo.
\newblock Size and form in efficient transportation networks.
\newblock {\em Nature}, 399:130--132, May 1999.

\bibitem{press92}
W.~H. Press, S.~A. Teukolsky, W.~T. Vetterling, and B.~P. Flannery.
\newblock {\em Numerical Recipes in C}.
\newblock Cambridge University Press, second edition, 1992.

\bibitem{heusner91a}
A.~Heusner.
\newblock Body mass, maintenance and basal metabolism in dogs.
\newblock {\em J. Nutr.}, 121(11 Suppl):S8--17, Nov. 1991.

\bibitem{jungers85}
W.~Jungers, editor.
\newblock {\em Size and scaling in primate biology}.
\newblock Advances in Primatology. Plenum Press, New York, 1985.

\bibitem{allman99}
J.~M. Allman.
\newblock {\em Evolving brains}.
\newblock Scientific American Library, New York, 1999.

\bibitem{bennett87note}
Following Bennett and Harvey~\cite{bennett87}, we take one sample for each
  species of bird selecting those with lowest mass-specific resting metabolic
  rate. Note that we also include organisms that Bennett and Harvey state were
  measured during their active cycle whereas Bennett and Harvey do not. The use
  of other selection criteria does not greatly affect the results we present
  here.

\bibitem{lasiewski67}
R.~C. Lasiewski and W.~R. Dawson.
\newblock A re-examination of the relation between standard metabolic rate and
  body weight in birds.
\newblock {\em Condor}, 69:13--23, 1967.

\bibitem{kendeigh77}
S.~C. Kendeigh, V.~R. Dol'nik, and V.~M. Gavrilov.
\newblock Avian energetics.
\newblock In J.~Pinowski and S.~Kendeigh, editors, {\em Granivorous birds in
  ecosystems}, pages 129--204, 363--73, Cambridge, 1977. Cambridge University
  Press.

\bibitem{prothero86b}
J.~Prothero.
\newblock Scaling of energy-metabolism in unicellular organisms---a reanalysis.
\newblock {\em Comp. Biochem. Physiol. A-Physiol.}, 83(2):243--248, 1986.

\bibitem{patterson92a}
M.~Patterson.
\newblock A mass-transfer explanation of metabolic scaling relations in some
  aquatic invertebrates and algae.
\newblock {\em Science}, 255(5050):1421--1423, 1992.

\bibitem{patterson92b}
M.~Patterson.
\newblock Correction.
\newblock {\em Science}, 256(5058):722--722, 1992.

\bibitem{kinnear67}
J.~Kinnear and G.~Brown.
\newblock Minimum heart rates of marsupials.
\newblock {\em Nature}, 215:1501, 1967.

\bibitem{dodds2000ua}
P.~S. Dodds and D.~H. Rothman.
\newblock Geometry of river networks {I}: {S}caling, fluctuations, and
  deviations.
\newblock Submitted to Phys. Rev. E, 2000.

\bibitem{degroot1975}
M.~H. DeGroot.
\newblock {\em Probability and Statistics}.
\newblock Addison-Wesley, Reading, Massachusetts, 1975.

\bibitem{hastie87}
T.~Hastie and R.~Tibshirani.
\newblock Generalized additive models: {S}ome applications.
\newblock {\em J. Amer. Stat. Assoc.}, 82:371--86, 1987.

\bibitem{cohn54a}
D.~Cohn.
\newblock Optimal systems {I}. {T}he vascular system.
\newblock {\em Bull. Math. Biophys.}, 16:59--74, 1954.

\bibitem{cohn54b}
D.~Cohn.
\newblock Optimal systems {I}{I}. {T}he vascular system.
\newblock {\em Bull. Math. Biophys.}, 17:219--227, 1955.

\bibitem{rashevsky62}
N.~Rashevsky.
\newblock General mathematical principles in biology.
\newblock In N.~Raschevsky, editor, {\em Physicomathematical Aspects of
  Biology}, Proceedings of the International School of Physics ``Enrico
  Fermi''; course 16, pages 493--524, New York, 1962. Academic Press.

\bibitem{mlabarbera90}
M.~LaBarbera.
\newblock Principles of design of fluid transport systems in zoology.
\newblock {\em Science}, 249:992--1000, 1990.

\bibitem{turcotte98}
D.~L. Turcotte, J.~D. Pelletier, and W.~I. Newman.
\newblock Networks with side branching in biology.
\newblock {\em J. Theor. Biol.}, 193:577--592, 1998.

\bibitem{lamb45}
H.~Lamb.
\newblock {\em Hydrodynamics}.
\newblock Dover, New York, 6th edition, 1945.

\bibitem{womersley55a}
J.~R. Womersley.
\newblock Method for the calculation of velocity, rate of flow and viscous drag
  in arteries when the pressure gradient is known.
\newblock {\em J. Physiol.}, 127:553--563, 1955.

\bibitem{murray26}
C.~D. Murray.
\newblock The physiological principle of minimum work. {I}. {T}he vascular
  system and the cost of blood volume.
\newblock {\em Proc. Natl. Acad. Sci. U.S.A}, 12:207--214, 1926.

\bibitem{womersley55b}
J.~R. Womersley.
\newblock Oscillatory motion of a viscous liquid in a thin-walled elastic
  tube---{I}: The linear approximation for long waves.
\newblock {\em Phil. Mag.}, 46(373):199--221, 1955.

\bibitem{fung90}
Y.-C.~B. Fung.
\newblock {\em Biomechanics: motion, flow, stress, and growth}.
\newblock Springer-Verlag, New York, 1990.

\bibitem{kassab93a}
G.~S. Kassab, C.~A. Rider, T.~N. J., and Y.~B. Fung.
\newblock Morphometry of pig coronary arterial trees.
\newblock {\em Am. J. Physiol.}, 265:H350--H365, 1993.

\bibitem{kassab93b}
K.~Kassab, G. S.~Imoto, F.~C. White, C.~A. Rider, Y.-C.~B. Fung, and C.~M.
  Bloor.
\newblock Coronary arterial tree remodeling in right ventricular hypertropy.
\newblock {\em Am. J. Physiol.}, 265:H366--H375, 1993.

\bibitem{zamir83}
M.~Zamir, S.~M. Wrigley, and B.~L. Langille.
\newblock Arterial bifurcations in the cardiovascular system of a rat.
\newblock {\em J. Gen. Physiol.}, 81:325--335, 1983.

\bibitem{hoppeler81}
H.~Hoppeler, O.~Mathieu, E.~Weibel, R.~Krauer, S.~Lindstedt, and C.~Taylor.
\newblock Design of mammalian respiratory system {V}{I}{I}{I}. {C}apillaries in
  skeletal muscles.
\newblock {\em Respir. Physiol.}, 44:129--150, 1981.

\bibitem{bishop99}
C.~M. Bishop.
\newblock The maximum oxygen consumption and aerobic scope of birds and
  mammals: getting to the heart of the matter.
\newblock {\em Proc. Roy. Lond. B.}, 266:2275--81, 1999.

\bibitem{stahl67}
W.~R. Stahl.
\newblock Scaling of respiratory variables in mammals.
\newblock {\em Journal of Applied Physiology}, 22:453--460, 1967.

\bibitem{speakman90}
J.~Speakman.
\newblock On {B}lum's four-dimensional geometric explanation for the 0.75
  exponent in metabolic allometry.
\newblock {\em J. Theor. Biol.}, 144(1):139--141, 1990.

\end{thebibliography}
\end{document}